\documentclass[hidelinks,sort&compress,3p,twoside,twocolumn]{article}

\usepackage[labelfont=bf]{caption}
\usepackage{tikz}
\usepackage{multirow}
\usepackage{natbib}
 \usepackage{mathrsfs}
\usepackage{amsmath,epsfig,graphicx}

\usepackage{subfigure}

\usepackage{color}
\usepackage[affil-it]{authblk}
\usepackage{wrapfig}

\DeclareMathOperator*{\argmax}{argmax}
\DeclareMathOperator*{\argmin}{argmin}

\usepackage{accents}
\newlength{\dhatheight}

\title{Maximum likelihood smoothing estimation in state-space models: \\An incomplete-information based approach}

\author[]{Budhi Arta Surya\footnote{Email address: \texttt{budhi.surya@vuw.ac.nz}}}
\affil[]{School of Mathematics and Statistics, Victoria University of Wellington, \\Gate 6, Kelburn PDE, Wellington 6140, New Zealand}

\usetikzlibrary{automata,positioning}

\allowdisplaybreaks

\bibpunct{(}{)}{;}{a}{,}{,}

\usepackage{amsfonts,amssymb,amsthm,booktabs,color,epsfig,graphicx,hyperref,url}

\theoremstyle{plain}% Theorem-like structures provided by amsthm.sty
\newtheorem{theorem}{Theorem}
\newtheorem{proposition}{Proposition}
\newtheorem{lemma}{Lemma}
\newtheorem{corollary}{Corollary}
\newtheorem{remark}{Remark}

\begin{document}

\maketitle \pagestyle{myheadings} \markboth{B.A. Surya}{Maximum likelihood recursive estimation in state-space models}

\begin{abstract}
This paper revisits the work of \cite{Rauch} and develops a novel method for recursive maximum likelihood state estimation in general state-space models. The new method is based on statistical analysis of incomplete observations of the systems. Score function and conditional observed information of the incomplete observations/data are introduced and their distributional properties are discussed. Some identities concerning the score function and information matrices of the incomplete data are derived. Maximum likelihood estimation of state-vector is presented in terms of the score function and observed information matrices. In particular, to deal with nonlinear state-space, a sequential Monte Carlo method is developed. It is given recursively by an EM-gradient-particle filtering which extends the work of \cite{Lange} for state estimation. To derive covariance matrix of state-estimation errors, an explicit form of observed Fisher information matrix is proposed. It extends \cite{Louis} general formula for the same matrix to state-vector estimation. Under (Neumann) boundary conditions of state transition probability distribution, the inverse of this matrix coincides with the Cram\'er-Rao lower bound on the covariance matrix of estimation errors of unbiased state-estimator. In the case of linear models, the method shows that the Kalman filter is a fully efficient unbiased state estimator whose covariance matrix of estimation error coincides with  the Cram\'er-Rao lower bound. Some numerical examples are discussed to exemplify the main  results. 
\medskip


%\noindent \textbf{MSC 2020}: 60J20; 60J27; 60J28; 62N99; 62H05

\end{abstract}

\section{Introduction}\label{sec:1}
The recursive filtering algorithm proposed by \cite{Kalman} has been an important tool for online estimation of (latent) state variables which are partially observed in noisy measurements. It has found immense applications in various different fields spanning from engineering for tracking and navigation (see e.g., \cite{Bar-Shalom}, \cite{Musoff2005}, \cite{Schmidt66} and \cite{Smith62}), economics/time series analysis (see e.g., \cite{Durbin}, \cite{Harvey}, \cite{Kitagawa2021}), molecular and RNA diagnoses (\cite{Kelemen}, \cite{Vaziri}), just to mention a few. See relevant literature therein for more details. The key to the development of the Kalman filter lies on the state-space representation of an underlying physical/economical/biological systems under consideration. See \cite{Anderson}, \cite{Bar-Shalom} and above references. 

In the derivation of the Kalman filtering, the latent state-vector is estimated using all observations up to recent time step. The derivation assumes absence of delay between the receipt of the last measurement and production of the state estimator. In the presence of delay, more observations become available during the delay period which can be used to produce a more accurate estimates of the state than the Kalman filter. The results are of particular significance in post-experimental data analysis to obtain a refined estimate of the state-vector. 

This is the smoothing estimation problem. Among the early references on the state-space approach to smoothing estimation for linear dynamical systems are those of \cite{Rauch63} and \cite{Rauch} who developed sequential algorithms for fixed-interval, fixed-point and fixed-lag smoothing for discrete time-step using maximum likelihood approach. \cite{Meditch67} extended the results of the latter two papers to a continuous-time linear state-space system based on a discrete-time skeleton of the continuous-time system. \cite{Bryson63,Bryson75} considered the fixed-interval smoothing problem for continuous-time, linear and nonlinear state-space system. The authors reformulated the smoothing estimation as a linear quadratic regulator problem, rather than a maximum likelihood estimation, for a continuous-time system which was solved using Bellman's principles of optimality and dynamic programming. The latter reduces the smoothing problem to that of solving the corresponding Bellman systems of parabolic differential equations, which for nonlinear systems in general is difficult to solve. See for e.g. \cite{Meditch73} for a comparative survey on this.

To deal with nonlinear state-space models, \cite{Psiaki} used localization method by linearizing the state-space around the a posteriori estimate of state to develop a recursive equation for smoothing, using similar idea to the extended Kalman filter (EKF).  See e.g. \cite{Jazwinski}, \cite{Anderson}, \cite{Bar-Shalom}, and \cite{Musoff2005} for further details on EKF. The method is called \textit{backward-smoothing EKF}. The localization method implies that the required probability distribution of the state-vector is approximated by a Gaussian distribution which causes a distortion of the true distribution of the state-vector resulting in estimation divergence. Hence, the backward-smoothing EKF is suboptimal. To overcome this difficulty, \cite{Hurzeler}, \cite{Doucet2000}, \cite{Godsill2004}, \cite{Vo} and \cite{Saha} developed a \textit{forward-backward smoothing} scheme based on a sequential Monte Carlo (SMC) (importance) sampling from conditional distribution $f(x_k\vert y_{0:n},\theta)$, with $y_{0:n}=\{y_0,\ldots,y_n\}$, $n\geq k\geq 0$, whereas $\theta$ is the parameter of the considered stochastic system. The key to developing the backward recursion is central to the use of smoothing distribution $f(x_k\vert y_{0:n},\theta\}$ developed earlier on in \cite{Kitagawa87}, whilst the forward equation involves particle filtering for computing a posteriori distribution $f(x_k\vert y_{0:k},\theta)$. The particle filtering itself has been successfully used in variety of applications in various fields. See \cite{Godsill} for a survey of recent developments of the particle filtering, see also \cite{Doucet,Doucet2000}, \cite{Kitagawa2021,Kitagawa,Kitagawa93}, \cite{Ristic}, \cite{Shephard} and \cite{Trianta}. Other SMC smoothing schemes such as \textit{filter-smoothing} was discussed in \cite{Kitagawa}, the \textit{two-filter smoothing} in \cite{Briers} and \cite{Fearnhead}, and the \textit{block-based smoothing} by \cite{Fong}. The latter is based on randomly sampling particles from $f(x_{1:n}\vert y_{1:n},\theta)$ developed based on earlier work of \cite{Tanizaki} in which the author proposed a method for generating random draws from the joint smoothing distribution $f(x_k, x_{k+1}\vert y_{1:n},\theta)$.

The merit of the above mentioned SMC smoothing schemes is that the numerical method is applicable for general state-space models by random sampling from the (joint) smoothing distribution to get smoothed estimate $\widehat{x}_{k\vert n}:=\mathbb{E}[x_k\vert y_{0:n},\theta]$ of the state-vector $x_k$. By iterated law of conditional expectation, one can show that $\widehat{x}_{k\vert n}$ has a minimum mean square error (MMSE). However, due to highly numerical nature of the scheme, the method lacks in providing an explicit solution to the smoothing problem, even in the linear state-space models, and in producing an estimate of the covariance matrix of estimation errors, which can be used to construct a $95\%$ confidence interval for  the state-vector $x_k$. 

Maximum likelihood (ML) based smoothing estimation was discussed (for linear Gaussian state-space model) by \cite{Rauch}. In this approach, smoothed estimates $\widehat{x}_{k\vert n}^s$ and $\widehat{x}_{k+1\vert n}^s$ of the states $x_k$ and $x_{k+1}$ are found as the maximizer of the logarithm of the joint smoothing distribution $f(x_k, x_{k+1}\vert y_{0:n},\theta)$. Equivalently, the smoother $\widehat{x}_{k\vert n}^s$ represents the value of $x_k$ at which the conditional distribution $f(x_k\vert \widehat{x}_{k+1\vert n}^s,y_{0:n},\theta)$ is maximal. Considering the filtering distribution $f(x_k\vert y_{0:k},\theta)=N(x_k\vert \widehat{x}_{k\vert k}, P_{k\vert k})$, with $P_{k\vert k}$ being the posteriori estimate of covariance matrix of estimation error, \cite{Rauch} were able to get an explicit solution $\widehat{x}_{k\vert n}^s$ to the smoothing problem. Under the same consideration, \cite{Vo} derived an explicit solution to the forward-backward smoothing for Gaussian mixture linear system. In the case that the smoothing distribution $f(x_k\vert y_{0:n},\theta)$ is unimodal and symmetry, the ML smoothing estimator $\widehat{x}_{k\vert n}^s$ and the MMSE $\widehat{x}_{k\vert n}$ coincide. Otherwise, they comprise two different unbiased smoothed estimators of the state-vector $x_k$. 

Although the normality of $f(x_k\vert y_{0:k},\theta)$ was not verified in \cite{Rauch}, it may not be necessary to use for deriving the ML smoother $\widehat{x}_{k\vert n}^s$ for linear state-space mdel. In general, it is not available explicitly, see for e.g. \cite{Godsill} and \cite{Kitagawa2021}. For nonlinear state-space, the ML smoothing estimation and the imposed condition on the smoothing distribution given in \cite{Rauch} may not be applicable to derive an explicit smoothed estimator of the state. To deal with general state-space system, \cite{Saha} proposed a SMC scheme based on the forward-backward smoothing. Following the argument, as used in \cite{Godsill2001} for particle filtering, \cite{Saha} found $\widehat{x}_{k\vert n}^s$ by applying a randomized adaptive grid approximation of the values of the smoothing distribution $f(x_k\vert y_{0:n},\theta)$.

The work of \cite{Rauch}, known as the \textit{Rauch-Tung-Striebel} (RTS) \textit{smoother} has been widely used and well cited in literature\footnote{\url{https://arc.aiaa.org/doi/pdfplus/10.2514/3.3166}}. In the context of parameter estimation, \cite{Liu2022} applied the RTS smoother for ML estimation of bilinear state-space system using expectation-maximization (EM) algorithm. \cite{Teymouri} used the EKF and EM algorithm for simultaneous estimation of state and parameter of continuous-time linear system. \cite{Ramadan} applied Bayes' formula repeatedly to smoothing distribution $f(x_k\vert x_{k+1},y_{0:n},\theta)$ and used EM algorithm to find the smoother $\widehat{x}_{k\vert n}$. In the M-step of the EM algorithm, an updated $\widehat{x}_{k\vert n}^{(\ell+1)}$ is specified by $\argmax_{x_k} \mathbb{E}[\log f(x_k\vert x_{k+1},y_{0:n},\theta)\vert \widehat{x}_{k\vert n}^{(\ell)},y_{0:n},\theta]$. However, due to nonlinearity of the smoothing distribution, $\widehat{x}_{k\vert n}^{(\ell+1)}$ is not explicit. The smoothing was then solved using \cite{Doucet2000} forward-backward algorithm. 

However, it is known in particular for parameter estimation that the EM-algorithm has (semi) linear rate of convergence compared to the EM-gradient algorithm of \cite{Lange}, which achieves its convergence at quadratic rate, hence faster than the EM-algorithm. See also \cite{McLachlan}. Beside that, as will be shown later, the EM algorithm may not be able to produce for nonlinear system an explicit update in the M-step of the algorithm. Furthermore, the EM-algorithm most notably does not directly produce the score function nor the observed information matrices of the observed data needed to derive the ML smoother and the covariance matrix, respectively. Although \cite{VanTrees68}, \cite{VanTrees}, \cite{Gill}, \cite{Tichavsky}, \cite{Bar-Shalom}, \cite{Bergman}, \cite{Ristic} and \cite{Zhao} provided Bayesian Cram\'er-Rao lower bounds for the covariance matrix of (random) parameter estimates, their results however are not directly applicable for the ML state estimation, see \cite{Surya2022b} for details. 

As far as the above references are concerned, there has been a gap in literature over discussions on the ML smoothing state-estimation and the corresponding covariance matrix of estimation error for general stochastic state-space systems. These concerns have been largely unexamined in literature since \cite{Rauch}.

\noindent \textbf{Research contributions}
%In summary following the above literature, maximum likelihood smoothing estimation can be classified into
%%
%\begin{enumerate}
%\item[(i)] Bayesian method and its variations based on \textit{incomplete-data likelihood function} of $(x_k,x_{k+1}, y_{0:n})$
%%
%\begin{align}
%&f(x_k,x_{k+1},y_{0:n}\vert \theta)=f(x_{k+1}\vert x_k,\theta)f(x_k\vert y_{0:k},\theta) \nonumber\\
% &\hspace{2cm}\times f(y_{k+1},\ldots,y_n\vert x_{k+1},\theta)f(y_{0:k}\vert\theta). \label{eq:eq1}
% \end{align}
% %
% The key to implementing the above Bayes formula is the knowledge of filtering distribution $f(x_k\vert y_{0:k},\theta)$, which is in general not available explicitly. For Gaussian linear system, \cite{Rauch} used $f(x_K\vert y_{0:k},\theta)=N(x_k\vert H_k \widehat{x}_{k\vert k}, P_{k\vert k})$, for a conformable measurement matrix $H_k$, to derive an explicit form of the ML smoother $\widehat{x}_{k\vert n}^s$.
% 
% \item[(ii)] Dynamic optimization of \textit{complete-data loglikelihood function} $\log f(x_{0:n},y_{0:n}\vert\theta)$ over state-vector $(x_0,\ldots,x_n)$. The problem is solved using Bellman's principle of optimality and dynamic programming.
% 
% \item[(iii)] A randomized adaptive grid approximation of the values of the smoothing distribution $f(x_k\vert y_{0:n},\theta)$.
% 
% \end{enumerate}

%From the above references, we may classify the existing filtering/smoothing method into two classes: least square with/out localization and Bayesian approach, see \cite{Trianta}. 
Motivated by recent work of \cite{Surya2022b,Surya2022}, this paper proposes a novel approach for recursive ML smoothing estimation of state. The novelty of the proposed approach is central to the incomplete-data formulation of a stochastic state-space systems under consideration. Within this framework, statistical analysis of incomplete data is used to solve the smoothing problem. To be more details, we consider $(x_k, x_{k+1}, y_{0:n})$, $k\leq n$, as an incomplete observation/data of the considered stochastic system represented by a complete data $(x_{0:n}, y_{0:n})$. 

The corresponding score function and conditional observed information matrices of the incomplete data are introduced. In particular, the \textit{incomplete-data score function} is obtained as an orthogonal projection of \textit{complete-data score function} onto Hilbert space spanned by the incomplete data. Recursive equation for $\widehat{x}_{k\vert n}^s$ is developed in terms of the score function and information matrices. It is given in terms of an EM-gradient-particle recursion, extending the work of \cite{Lange} for smoothing. The recursion is shown to be locally equivalent to an EM smoothing algorithm. As opposed to the latter, the recursion has an explicit iteration update. 

The results show that the ML smoother $\widehat{x}_{k\vert n}^s$ gives an optimal estimate of unknown state $x_k^0$ with more adherence of loglikehood having less covariance matrix than that of the ML estimator $\widehat{x}_k$. To derive estimated standard error of $\widehat{x}_{k\vert n}^s$, an iterative backward equation for covariance matrix of estimation error $\widehat{x}_{k\vert n}^s-x_k$ is given. Their availability in general form makes the method available for ML smoothing estimation in general state-space models. For the case of Gaussian linear system, the method reproduces the results of \cite{Rauch}. Furthermore, the method shows that the RTS smoother is a fully efficient ML estimator whose covariance matrix equals to the inverse of expected information matrix.

This paper is organized as follows. Section \ref{sec:sec2} discusses some results on finite-sample properties of incomplete-data. The results are used in Section \ref{sec:sec3} for solving ML smoothing problem. Applications to linear and nonlinear state-space are discussed in Section \ref {sec:sec4}. Section \ref{sec:numeric} presents numerical examples before the paper is concluded in Section \ref{sec:conclusion}. Lengthy proofs are deferred to Appendix.

\section{Finite-sample properties of incomplete data}\label{sec:sec2}
In order to formulate ML smoothing estimation from incomplete data, we consider the random vector $(x_k, x_{k+1},y_{0:n})$, $k\leq n-1$, $n\geq 1$, as a post-experimental incomplete observations/data of realized values $(x_{0:n},y_{0:n})$ of a stochastic system under consideration. The section below discusses some statistical properties of incomplete data in terms of likelihood function, score function and observed information matrices. They are developed based on \cite{Surya2022b,Surya2022}.

\subsection{Likelihood function of incomplete data}
Let $z$ and $\xi$ be two random vectors defined on the same probability space $(\Omega,\mathscr{F},\mathbb{P})$. Denote by $(\mathcal{Z},\mathcal{S})$ and $(\Xi,\mathcal{T})$ the corresponding measurable state spaces of $z$ and $\xi$ and by $T:\mathcal{Z}\rightarrow \Xi$ a many-to-one mapping from $\mathcal{Z}$ to $\Xi$. Suppose that a \textit{complete-data} vector $z\in\mathcal{Z}$ is partially observed through an \textit{incomplete-data} vector $\xi=T(z)$ in $\Xi$. Furthermore, assume that there exists probability density functions $f(z\vert\theta)$ and $f(\xi\vert\theta)$ corresponding to the complete data-vector $z\in\mathcal{Z}$ and its incomplete observation $\xi\in\Xi$, respectively. Here $\theta$ represents a vector of parameters on a parameter space $\Theta$ characterizing the probability distribution of $z$. Define $\mathcal{Z}(\xi)=\{z\in\mathcal{Z}: T(z)=\xi\}\in \mathcal{S}$. Then, it follows from \cite{Halmos} that $f(\xi\vert \theta)$ is given by
\begin{align}\label{eq:halmos}
f(\xi\vert \theta)=\int_{\mathcal{Z}(\xi)} f(z\vert\theta)\lambda(dz),
\end{align}
where $\lambda$ is a $\sigma-$finite measure on $\mathcal{S}$, absolutely continuous w.r.t the distribution $\mathbb{P}\circ z^{-1}$ with probability density function $f(z\vert \theta)$ (the Radon-Nikodym derivative). 
%Notice that our description of the marginal distribution (\ref{eq:halmos}) is slightly more general than the one used in McLachlan and Krishnan (2008).

%Following identity (\ref{eq:halmos}), the conditional probability density function $f(z\vert \xi,\theta)$ of the complete-data $z$ given its incomplete observation $\xi$ is therefore specified by
%%
%\begin{align}\label{eq:condpdf}
%f(z\vert \xi,\theta)=\frac{f(z\vert\theta)}{f(\xi\vert \theta)}.
%\end{align}
Without loss of generality, we assume that the log-likelihood function $\log f(z\vert\theta)$ is twice continuously differentiable w.r.t $\theta$ and for all $\theta\in\Theta$, $m\in\{0,1,2\},$
\begin{align}\label{eq:bounded}
\int_{\mathcal{Z}(\xi)} \Big\vert \frac{\partial^m \log f(z\vert\theta) }{\partial \theta^m}\Big\vert f(z\vert \xi,\theta)\lambda(dz)<\infty. \tag{A1}
\end{align}
Condition (\ref{eq:bounded}) verifies the existence of expectation $\mathbb{E}\big[\big\vert \frac{\partial^m \log f(z\vert\theta) }{\partial \theta^m}  \big\vert \big\vert \xi,\theta\big]$ for all $\theta\in\Theta$ and $m\in\{0,1,2\}$. 

In the context of smoothing, the random vectors 
\begin{align*}
\xi:=(x_k,x_{k+1},y_{0:n}) \;\; \textrm{and} \;\; z:=(x_{0:n},y_{0:n})
\end{align*}
respectively represents the post-experimental incomplete and complete observations of realized values of a stochastic state-space system under consideration. 

From the view point of ML smoothing estimation, the smoother $\widehat{x}_{k\vert n}^s$ is the value of $x_k$ that maximizes the loglikelihood function $\log f(x_k,y_{0:n}\vert\theta)$. Similarly, $\widehat{x}_{k\vert n}^s$ and $\widehat{x}_{k+1\vert n}^s$ are respectively the values of $x_k$ and $x_{k+1}$ that maximize $\log f(x_k,x_{k+1},y_{0:n}\vert\theta)$. That is to say,
\begin{align}\label{eq:maximizer}
(\widehat{x}_{k\vert n}^s,\widehat{x}_{k+1\vert n}^s)=\argmax_{x_k,x_{k+1}}\log f(x_k,x_{k+1},y_{0:n}\vert\theta).
\end{align} 
Equivalently, in terms of the score function, $\widehat{x}_{k\vert n}^s$ solves
%%
%\begin{align*}
%0=\frac{\partial \log f(\xi\vert \theta)}{\partial x_k}=\frac{\frac{\partial }{\partial x_k} \int_{\mathcal{Z}(\xi)}f(z\vert \theta) d \lambda(z)}{\int_{\mathcal{Z}(\xi)}f(z\vert \theta) d \lambda(z)},
%\end{align*}
%%%
\begin{align*}
0=\frac{\partial \log f(\xi\vert \theta)}{\partial x_k}=\frac{\partial }{\partial x_k} \log \int_{\mathcal{Z}(\xi)}f(z\vert \theta) d \lambda(z),
\end{align*}
similarly defined for $\widehat{x}_{k+1\vert n}^s$. The difficulty in evaluating integral (\ref{eq:halmos}) arises due to attributed complexity of the likelihood function $f(z\vert \theta)$ of the complete data $z$. Meanwhile, the Bayes formula derived in \cite{Rauch},
\begin{align}
&f(x_k,x_{k+1},y_{0:n}\vert \theta)=f(x_{k+1}\vert x_k,\theta)f(x_k\vert y_{0:k},\theta) \nonumber\\
 &\hspace{2cm}\times f(y_{k+1},\ldots,y_n\vert x_{k+1},\theta)f(y_{0:k}\vert\theta), \label{eq:eq1}
 \end{align}
requires the knowledge of filtering distribution $f(x_k\vert y_{0:k},\theta)$ which is in general not available in closed form. Hence, this posts difficulty in deriving explicit $\widehat{x}_{k\vert n}^s$ and $\widehat{x}_{k+1\vert n}^s$. 
 
 \subsection{Score functions and conditional observed information matrices of the state-vector $(x_k,x_{k+1})$}

To overcome the smoothing problem, an explicit solution is proposed in terms of the score functions and observed information matrices of incomplete data, which are discussed in further details in the section below. 

\subsubsection{Score functions of the state-vector $(x_k,x_{k+1})$}
For this purpose, let $\eta=x_{0:n}/\{x_k,x_{k+1}\}$ be a vector consisting of all states in $x_{0:n}$, except $(x_k,x_{k+1})$, and denote by $\mathcal{H}$ the space of all possible values of $\eta$. Furthermore, notice that the marginal distribution (\ref{eq:halmos}) of the random vector $\xi$ can be equivalently rewritten as
\begin{align}\label{eq:int}
f(\xi\vert \theta)=\int_{\mathcal{H}} f(\xi,\eta\vert \theta) d\lambda(\eta),
\end{align}
which leads to the following identity for the conditional probability density functions $f(z\vert \xi,\theta)$ and $f(\eta \vert \xi,\theta)$,
\begin{align}\label{eq:identity}
f(z\vert \xi,\theta)=\frac{f(z\vert \theta)}{f(\xi\vert \theta)}=f(\eta\vert \xi,\theta).
\end{align}

Taking logarithm and derivative with respect to $x_k$ on both sides of (\ref{eq:identity}) yields the relationship between the score functions of complete $z$ and incomplete data $\xi$:
\begin{align}\label{eq:eq3}
\frac{\partial \log f(\xi\vert \theta)}{\partial x_k}=\frac{\partial \log f(z\vert \theta)}{\partial x_k} - \frac{\partial \log f(z\vert \xi,\theta)}{\partial x_k}.
\end{align}
\begin{lemma}\label{lem:lem1}
For any given $\theta$ and observation $\xi$,
\begin{align*}
\mathbb{E}\big[\frac{\partial \log f(z\vert \xi,\theta)}{\partial x_k}\big\vert \xi,\theta\big]=0.
\end{align*}
\end{lemma}
\noindent \textbf{Proof.} By Bayes formula and dominated convergence, $\mathbb{E}\big[\frac{\partial \log f(z \vert \xi,\theta)}{\partial x_k}\big\vert \xi,\theta\big]
=\int_{\mathcal{Z}(\xi)} \frac{\partial \log f(z\vert \xi,\theta)}{\partial x_k}  f(z\vert \xi,\theta) d\lambda (z)\\ = \int_{\mathcal{H}} \frac{\partial \log f(z\vert \xi,\theta)}{\partial x_k}  f(\eta \vert \xi,\theta) d\lambda (\eta)=\frac{\partial}{\partial x_k}\int_{\mathcal{H}} f(\eta\vert \xi,\theta) d\lambda (\eta),$
which by (\ref{eq:int})-(\ref{eq:identity}) completes the proof of the claim. $\square$

Taking conditional expectation $\mathbb{E}[\bullet \vert \xi_k,\theta]$ on both sides of (\ref{eq:eq3}), we obtain an identity for the score function.

\begin{proposition}\label{prop:main}
For any given $\theta$ and observation $\xi$,
\begin{align}\label{eq:mainidentity}
\frac{\partial \log f(\xi\vert \theta)}{\partial x_k}=\mathbb{E}\big[\frac{\partial \log f(z\vert \theta)}{\partial x_k}\big\vert \xi,\theta\big].
\end{align}
\end{proposition}

Identity (\ref{eq:mainidentity}) also holds for the score function of $x_{k+1}$.

\subsubsection{Observed information matrices}
To derive the ML smoother and the standard error, we introduce the following (conditional) observed information matrices. First, consider observed information matrix $J^{\xi}(\xi\vert\theta)$ defined by the block partitioned matrix
\begin{align*}
\mathbf{J}^{\xi}(\xi\vert\theta)=\left(
\begin{array}{cc}
J_{x_k,x_k}^{\xi}(\xi\vert\theta) & J_{x_k,x_{k+1}}^{\xi}(\xi\vert\theta)\\
J_{x_{k+1},x_k}^{\xi}(\xi\vert\theta) & J_{x_{k+1},x_{k+1}}^{\xi}(\xi\vert\theta)
\end{array}\right),
\end{align*}
where the first column vector of the matrix is given by
\begin{align*}
J_{x_k,x_k}^{\xi}(\xi\vert\theta)\equiv& - \frac{\partial^2 \log f(\xi\vert \theta)}{\partial x_k \partial x_k^{\top}},\\[2pt]
J_{x_{k+1},x_k}^{\xi}(\xi\vert\theta)\equiv& - \frac{\partial^2 \log f(\xi\vert \theta)}{\partial x_{k+1} \partial x_k^{\top}},
\end{align*}
whilst the second column vector is defined similarly. Secondly, we consider conditional observed information matrix $\mathbf{J}^z(\xi\vert\theta)$ whose first column vector is defined by
\begin{align*}
J_{x_k,x_k}^z(\xi\vert \theta)\equiv & \mathbb{E}\big[- \frac{\partial^2 \log f(z\vert \theta)}{\partial x_k \partial x_k^{\top}} \big\vert \xi,\theta\big],\\[2pt]
J_{x_{k+1},x_k}^z(\xi\vert \theta)\equiv & \mathbb{E}\big[- \frac{\partial^2 \log f(z\vert \theta)}{\partial x_{k+1} \partial x_k^{\top}} \big\vert \xi,\theta\big].
\end{align*}
Similarly defined for the second column vector. The third, denote by $\mathbf{J}^{z\vert \xi}(\xi\vert\theta)$ a conditional observed information matrix whose first column vector is specified by
\begin{align*}
J_{x_k,x_k}^{z\vert \xi}(\xi \vert\theta)\equiv & \mathbb{E}\big[- \frac{\partial^2 \log f(z\vert \xi, \theta)}{\partial x_k \partial x_k^{\top}} \big\vert \xi,\theta\big],\\[2pt]
J_{x_{k+1},x_k}^{z\vert \xi}(\xi \vert \theta)\equiv & \mathbb{E}\big[- \frac{\partial^2 \log f(z\vert \xi, \theta)}{\partial x_{k+1} \partial x_k^{\top}} \big\vert \xi,\theta\big],
\end{align*}
similarly defined for the second column vector. It follows from (\ref{eq:eq3}) that $\mathbf{J}^{\xi}(\xi\vert\theta)$, $\mathbf{J}^z(\xi\vert\theta)$ and $\mathbf{J}^{z\vert\xi}(\xi\vert\theta)$ satisfy
\begin{align*}
\mathbf{J}^{z\vert \xi}(\xi \vert\theta)= \mathbf{J}^z(\xi\vert \theta) - \mathbf{J}^{\xi}(\xi\vert \theta).
\end{align*}
\begin{proposition}
For any given $\theta$ and observation $\xi$,
\begin{align}\label{eq:positive}
\mathbf{J}^z(\xi\vert \theta) > \mathbf{J}^{\xi}(\xi\vert \theta).
\end{align}
%
%In particular, at $(\widehat{x}_{k\vert n}^s, \widehat{x}_{k+1\vert n}^s):=\argmax_{x_k,x_{k+1}} \log f(\xi\vert \theta)$,
%%
%\begin{align*}
%J^z(\xi\vert \theta) > J^{\xi}(\xi\vert \theta).
%\end{align*}
\end{proposition}
\noindent \textbf{Proof.} By symmetry property of the information matrix $\mathbf{J}^{z\vert \xi}(\xi \vert\theta)$, the proof is complete by showing that $\mathbf{J}^{z\vert\xi}(\xi\vert \theta)$ is positive definite which amounts to showing that $\mathbb{E}\big[-\frac{\partial^2 \log f(z\vert\xi,\theta)}{\partial x_{k} \partial x_k^{\top}}\big\vert \xi,\theta\big]=\mathbb{E}\big[\frac{\partial \log f(z\vert\xi,\theta)}{\partial x_{k}} \frac{\partial \log f(z\vert\xi,\theta)}{\partial x_{k}^{\top}}\big\vert \xi,\theta\big],$ which is the first diagonal element of $\mathbf{J}^{z\vert \xi}(\xi \vert \theta)$, and
\begin{align}\label{eq:eq9}
&\mathbb{E}\big[-\frac{\partial^2 \log f(z\vert\xi,\theta)}{\partial x_{k+1} \partial x_k^{\top}}\big\vert \xi,\theta\big] \nonumber\\
&\hspace{1.5cm}=\mathbb{E}\big[\frac{\partial \log f(z\vert\xi,\theta)}{\partial x_{k+1}} \frac{\partial \log f(z\vert\xi,\theta)}{\partial x_{k}^{\top}}\big\vert \xi,\theta\big],
\end{align}
the first off-diagonal element of $\mathbf{J}^{z\vert \xi}(\xi \vert \theta)$. The other two elements of $\mathbf{J}^{z\vert \xi}(\xi \vert \theta)$ follow similarly. The first identity was established in \cite{Surya2022b}, while the later can be proved in similar way to the former using the following identity (obtained by applying chain rule of derivative), $-\frac{\partial^2\log f(z\vert \xi,\theta)}{\partial x_{k+1} \partial x_k^{\top}} f(z\vert \xi,\theta)=-\frac{\partial }{\partial x_{k+1}}\big[\frac{\partial \log f(z\vert \xi,\theta)}{\partial x_k^{\top}} f(z\vert \xi,\theta)\big]+\frac{\partial \log f(z\vert \xi,\theta)}{\partial x_{k+1}}\frac{\partial \log f(z\vert \xi,\theta)}{\partial x_k^{\top}}.$
%
%\begin{align*}
%&\hspace{-0.5cm}-\frac{\partial^2\log f(z\vert \xi,\theta)}{\partial x_{k+1} \partial x_k^{\top}} f(z\vert \xi,\theta)=-\frac{\partial }{\partial x_{k+1}}\big[\frac{\partial \log f(z\vert \xi,\theta)}{\partial x_k^{\top}} f(z\vert \xi,\theta)\big]\\
%&\hspace{3cm}+\frac{\partial \log f(z\vert \xi,\theta)}{\partial x_{k+1}}\frac{\partial \log f(z\vert \xi,\theta)}{\partial x_k^{\top}}.
%\end{align*}
%%
Integrating both sides of the identity over the set $\mathcal{Z}(\xi)$ with respect to measure $d\lambda(z)$ establishes the claim. $\square$
\begin{corollary}
Deduced from (\ref{eq:positive}), for any given $\theta$ and $\xi$,
\begin{align}\label{eq:eq10}
J_{x_k,x_k}^z(\xi\vert\theta) > J_{x_k,x_k}^{\xi}(\xi\vert\theta).
\end{align}
\end{corollary}
Inequalities (\ref{eq:positive}) and (\ref{eq:eq10}) correspond to the resulting loss-of-information in finite-sample incomplete data, discussed in \cite{Surya2022b}, extending similar matrix inequality $\mathbf{I}^z(\theta)>\mathbf{I}^{\xi}(\theta)$ presented in \cite{Orchard} and Theorem 2.86 in \cite{Schervish} for expected information matrices (of parameter estimates)
\begin{align*}
\mathbf{I}^z(\theta):=\mathbb{E}\big[\mathbf{J}^z(\xi\vert\theta)\big\vert \theta\big] \; \textrm{and} \; 
\mathbf{I}^{\xi}(\theta):=\mathbb{E}\big[\mathbf{J}^{\xi}(\xi\vert\theta)\big\vert \theta\big].
\end{align*}
Unlike $\mathbf{J}^z(\xi\vert \theta)$ and $\mathbf{J}^{z\vert\xi}(\xi \vert \theta)$, the elements of information matrix $\mathbf{J}^{\xi}(\xi\vert\theta)$ are not linked to the complete-data loglikelihood function $\log f(z\vert \theta)$. Hence, they are difficult to evaluate. The results below overcome this difficulty.

\begin{theorem}\label{theo:theo2}
For given $\theta$ and observation $\xi$, $J_{x_k,x_k}^{\xi}(\xi\vert \theta)$ and $J_{x_{k+1},x_k}^{\xi}(\xi \vert \theta)$ are respectively defined by
\begin{align}
J_{x_k,x_k}^{\xi}(\xi\vert \theta)=& \mathbb{E}\big[- \frac{\partial^2 \log f(z\vert \theta)}{\partial x_k \partial x_k^{\top}} \big\vert \xi,\theta\big] \nonumber\\
&\hspace{-1.5cm}-\mathbb{E}\big[\frac{\partial \log f(z \vert \theta)}{\partial x_k} \frac{\partial \log f(z \vert \theta)}{\partial x_k^{\top}}    \big\vert \xi,\theta \big] \label{eq:louis1} \\
&\hspace{-1.5cm}+\mathbb{E}\big[\frac{\partial \log f(z \vert \theta)}{\partial x_k} \big\vert \xi,\theta \big]\mathbb{E}\big[\frac{\partial \log f(z \vert \theta)}{\partial x_k^{\top}} \big\vert \xi,\theta \big], \nonumber\\[4pt]
J_{x_{k+1},x_k}^{\xi}(\xi\vert \theta)=& \mathbb{E}\big[- \frac{\partial^2 \log f(z\vert \theta)}{\partial x_{k+1} \partial x_k^{\top}} \big\vert \xi,\theta\big] \nonumber\\
&\hspace{-1.5cm}-\mathbb{E}\big[\frac{\partial \log f(z \vert \theta)}{\partial x_{k+1}} \frac{\partial \log f(z \vert \theta)}{\partial x_k^{\top}}    \big\vert \xi,\theta \big] \label{eq:louis2}\\
&\hspace{-1.5cm}+\mathbb{E}\big[\frac{\partial \log f(z \vert \theta)}{\partial x_{k+1}} \big\vert \xi,\theta \big]\mathbb{E}\big[\frac{\partial \log f(z \vert \theta)}{\partial x_k^{\top}} \big\vert \xi,\theta \big]. \nonumber
\end{align}

\end{theorem}
\noindent \textbf{Proof.}
The first identity $J_{x_k,x_k}^{\xi}(\xi\vert \theta)$ was established in \cite{Surya2022b}. The second identity can be derived using similar arguments. See Appendix A for details. $\square$

Note that the information matrix $\mathbf{J}^{\xi}(\xi\vert \theta)$ extends the general matrix formula of \cite{Louis} for covariance matrix of parameter estimation to that of state vector.

%\subsection{Inverse of observed information matrices}

\section{Maximum likelihood smoothing estimation}\label{sec:sec3}
We further assume throughout the remaining of this paper that the state transition probability density function $f(x_k\vert x_{k-1}, \theta)$ has the following properties:
\begin{align}\label{eq:ass3}
f(x_k\vert x_{k-1},\theta)=0 \; \textrm{and} \; \frac{\partial f(x_k\vert x_{k-1},\theta)}{\partial x_k} =0, \tag{A2a}
\end{align}
for $x_k\in \partial \mathcal{X}_k$. Necessarily, for each $x_k\in \mathcal{X}_k$ and $n=0,1$,
\begin{align}\label{eq:ass3b}
\sup_{x_{k-1}}\Big\vert \frac{\partial^n f(x_k\vert x_{k-1},\theta)}{\partial x_k^n}\Big\vert <\infty, \tag{A2b}
\end{align}
where $\mathcal{X}_k$ denotes a set of all possible values of $x_k$. In fact, the above condition can as well be specified in terms of the prior distribution $f(x_k\vert \theta)$, see \cite{Surya2022b}. 

In the sequel below we denote by $\mathcal{S}(x_k,x_{k+1}\vert\theta)=\big(\mathcal{S}_{x_k}(x_k,x_{k+1}\vert\theta),\mathcal{S}_{x_{k+1}}(x_k,x_{k+1}\vert\theta)\big)^{\top}$ a column vector of respective state-vector $x_k$ and $x_{k+1}$ with
\begin{align*}
\mathcal{S}_{x_k}(x_k,x_{k+1}\vert\theta)=&\frac{\partial \log f(x_k,x_{k+1},y_{0:n}\vert\theta)}{\partial x_k},\\
\mathcal{S}_{x_{k+1}}(x_k,x_{k+1}\vert\theta)=&\frac{\partial \log f(x_k,x_{k+1},y_{0:n}\vert\theta)}{\partial x_{k+1}}.
\end{align*}
\begin{proposition}\label{prop:prop2}
Under (A2), it holds for any $\theta$ that
\begin{align}
\mathbb{E}\big[\mathcal{S}(x_k,x_{k+1}\vert\theta)\big\vert\theta\big]=&\mathbf{0},\label{eq:unbiased}\\
\mathbb{E}\big[\mathcal{S}(x_k,x_{k+1}\vert\theta)\mathcal{S}^{\top}(x_k,x_{k+1}\vert\theta)\big\vert\theta\big]=&\mathbf{I}^{\xi}(\theta), \label{eq:eq14}
\end{align}
with expected information matrix $\mathbf{I}^{\xi}(\theta):=\mathbb{E}[\mathbf{J}^{\xi}(\xi\vert\theta)\vert\theta]$.
%%
%\begin{align*}
%\mathbf{I}^{\xi}(\theta)=\left(
%\begin{array}{cc}
%\mathbb{E}[J_{x_k,x_k}^{\xi}(\xi\vert\theta)\vert\theta] & \mathbb{E}[J_{x_k,x_{k+1}}^{\xi}(\xi\vert\theta)\vert\theta]\\
%\mathbb{E}[J_{x_{k+1},x_k}^{\xi}(\xi\vert\theta)] & \mathbb{E}[J_{x_{k+1},x_{k+1}}^{\xi}(\xi\vert\theta)]
%\end{array}\right).
%\end{align*}
%% 
\end{proposition}
\noindent \textbf{Proof.}
See Appendix B for details of derivation. $\square$

It is straightforward to see from (\ref{eq:eq14}) that the information matrix $\mathbf{I}^{\xi}(\theta)$ is positive definite, hence invertible. The above identity leads to the following result.

\begin{proposition}
The smoother $(\widehat{x}_{k\vert n}^s,\widehat{x}_{k+1\vert n}^s)$ satisfies 
\begin{eqnarray}\label{eq:eq15}
\left(
\begin{array}{c}
\mathcal{S}_{x_k}(x_k,x_{k+1}\vert\theta)\\
\mathcal{S}_{x_{k+1}}(x_k,x_{k+1}\vert\theta)
\end{array}\right)
=\mathbf{I}^{\xi}(\theta)
\left(
\begin{array}{c}
\widehat{x}_{k\vert n}^s -x_k\\
\widehat{x}_{k+1\vert n}^s -x_{k+1}
\end{array}\right),
\end{eqnarray}
and is unbiased with the covariance matrix $\Sigma^s(\theta)=[\mathbf{I}^{\xi}]^{-1}(\theta)$. In particular, $\Sigma_{x_k,x_k}^s(\theta)$ is given recursively by
\begin{align}\label{eq:eq16}
&\Sigma_{x_k,x_k}^s(\theta)=\big(\mathbb{E}[J_{x_k,x_k}^{\xi}(\xi\vert\theta)\vert\theta]\big)^{-1} \mathbb{E}[J_{x_k,x_{k+1}}^{\xi}(\xi\vert\theta)\vert\theta] \nonumber\\
&\times\Sigma_{x_{k+1},x_{k+1}}^s(\theta) \mathbb{E}[J_{x_{k+1},x_k}^{\xi}(\xi\vert\theta)\vert\theta]\big(\mathbb{E}[J_{x_k,x_k}^{\xi}(\xi\vert\theta)\vert\theta]\big)^{-1} \nonumber\\
&+\big(\mathbb{E}[J_{x_k,x_k}^{\xi}(\xi\vert\theta)\vert\theta]\big)^{-1}.
\end{align}
\end{proposition}
\noindent \textbf{Proof.}
The identity (\ref{eq:eq15}) may be verified in part by looking at the state-vector $(x_k,x_{k+1})<(\widehat{x}_{k\vert n}^s,\widehat{x}_{k+1\vert n}^s)$ elementwise by which the identity gives the score vector $\mathcal{S}(x_k,x_{k+1}\vert\theta)>0$ elementwise. Vice versa, $\mathcal{S}(x_k,x_{k+1}\vert\theta)<0$ elementwise for $(x_k,x_{k+1})>(\widehat{x}_{k\vert n}^s,\widehat{x}_{k+1\vert n}^s)$. At $(x_k,x_{k+1})=(\widehat{x}_{k\vert n}^s,\widehat{x}_{k+1\vert n}^s)$, positive definiteness of $\mathbf{I}^{\xi}(\theta)$ implies that $\mathcal{S}(\widehat{x}_{k\vert n}^s,\widehat{x}_{k+1\vert n}^s)=\mathbf{0}$. These lead to the conclusion that $(\widehat{x}_{k\vert n}^s,\widehat{x}_{k+1\vert n}^s)$ is the maximizer (\ref{eq:maximizer}). Unbiasedness follows from (\ref{eq:unbiased}). By (\ref{eq:eq14}) the covariance matrix of $(\widehat{x}_{k\vert n}^s -x_k,\widehat{x}_{k+1\vert n}^s-x_{k+1})$ is specified by the inverse $[\mathbf{I}^{\xi}]^{-1}(\theta)$\footnote{following Theorem 7.2.1 on p.438 of \cite{Horn}, also Lemma 1.3, p.4 in \cite{Chui}.}, which coincides with the Cram\'er-Rao lower bound. The latter is due to $(\widehat{x}_{k\vert n}^s -x_k,\widehat{x}_{k+1\vert n}^s-x_{k+1})^{\top}$ given by a multiplicative constant matrix $[\mathbf{I}^{\xi}]^{-1}(\theta)$ of the vector score function. See e.g. \cite{Surya2022b} and \cite{VanTrees}. Applying inversion rule for block partitioned matrix, see e.g. \cite{Horn}, $\Sigma_{x_k,x_k}^s(\theta)$ is obtained. $\square$

\subsection{Recursive ML smoothing estimation}
Recall that (\ref{eq:maximizer}) yields smoothers $\widehat{x}_{k\vert n}^s$ and $\widehat{x}_{k+1\vert n}^s$ simultaneously. To reduce the complexity of smoothing problem (\ref{eq:maximizer}), $\widehat{x}_{k\vert n}^s$ can be found iteratively by considering 
\begin{equation}\label{eq:maximizer2}
\begin{split}
\widehat{x}_{k\vert n}^s=&\argmax_{x_k} \log f(x_k, \widehat{x}_{k+1\vert n}^s, y_{0:n}\vert \theta),\\
\widehat{x}_{n\vert n}^s=&\widehat{x}_{n\vert n},
\end{split}
\end{equation}
where $\widehat{x}_{n\vert n}=\argmax_{x_n} \log f(x_n,y_{0:n}\vert \theta)$ denotes the ML state estimator. The smoothing criterion (\ref{eq:maximizer2}) generalizes that of (3.27) considered in \cite{Rauch}. In this setting, the smoother $\widehat{x}_{k\vert n}^s$ solves the score equation
\begin{align}\label{eq:eq20}
%\frac{\partial \log f(x_k,\widehat{x}_{k+1\vert n}^s,y_{0:n}\vert \theta)}{\partial x_k}=0.
\mathcal{S}_{x_k}(x_k,\widehat{x}_{k+1\vert n}^s\vert\theta)=0.
\end{align}
\begin{theorem}\label{theo:maintheorem}
The ML smoother $\widehat{x}_{k\vert n}^s$ (\ref{eq:maximizer2}) satisfies 
\begin{align}\label{eq:eq21}
\widehat{x}_{k\vert n}^s =x_k + \mathcal{I}_{\xi}^{-1}(\theta)\mathcal{S}_{x_k}(x_k,\widehat{x}_{k+1\vert n}^s\vert \theta).
\end{align}
where $\mathcal{I}_{\xi}(\theta):=\mathbb{E}\big[-\frac{\partial^2 \log f(x_k,\widehat{x}_{k+1\vert n}^s,y_{0:n}\vert\theta)}{\partial x_k \partial x_k^{\top}}\big\vert\theta\big]$. Furthermore, the covariance matrix $\Sigma_{x_k,x_k}^s(\theta)$ coincides with 
\begin{equation}\label{eq:eq22}
\begin{split}
\Sigma_{x_k,x_k}^s(\theta)=&\mathcal{I}_{\xi}^{-1}(\theta)\mathbb{E}\big[\mathcal{S}_{x_k}(x_k,\widehat{x}_{k+1\vert n}^s \vert \theta)\\
&\times \mathcal{S}_{x_k}^{\top}(x_k,\widehat{x}_{k+1\vert n}^s\vert \theta)\big\vert\theta\big]\mathcal{I}_{\xi}^{-1}(\theta).
\end{split}
\end{equation}
\end{theorem}

It will be shown in Section \ref{sec:sec4} that (\ref{eq:eq21})-(\ref{eq:eq22}) (and (\ref{eq:eq16})) produce the RTS smoother of \cite{Rauch} and the covariance matrix of estimation error $\widehat{x}_{k\vert n}^s-x_k$.

\noindent \textbf{Proof.}
Using similar verification arguments to (\ref{eq:eq15}), the ML smoother $\widehat{x}_{k\vert n}^s$ satisfies $\mathcal{S}_{x_k}(x_k,\widehat{x}_{k+1\vert n}^s\vert \theta)=\mathcal{I}_{\xi}(\theta)\big(\widehat{x}_{k\vert n}^s -x_k\big)$. To be more precise, $\mathcal{S}_{x_k}(x_k,\widehat{x}_{k+1\vert n}^s\vert \theta)\gtrless0$ for $x_k\lessgtr\widehat{x}_{k\vert n}^s$, and $\mathcal{S}_{x_k}(x_k,\widehat{x}_{k+1\vert n}^s\vert \theta)=0$ for $x_k=\widehat{x}_{k\vert n}^s$ which in turn leads to $\widehat{x}_{k\vert n}^s$ being the maximizer (\ref{eq:maximizer2}). By positive definiteness of $\mathcal{I}_{\xi}(\theta)$, the expression (\ref{eq:eq21}) of $\widehat{x}_{k\vert n}^s$ is obtained by multiplying the former equation by the inverse $\mathcal{I}_{\xi}^{-1}(\theta)$. By equivalence of the two criterions (\ref{eq:maximizer}) and (\ref{eq:maximizer2}), the covariance matrix $\Sigma_{x_k,x_k}(\theta)$ is derived directly from $\mathbb{E}\big[(\widehat{x}_{k\vert n}^s - x_k)(\widehat{x}_{k\vert n}^s - x_k)^{\top}\big\vert \theta\big]$. $\square$
\begin{remark} Since in the score function $\mathcal{S}_{x_k}(x_k,\widehat{x}_{k+1\vert n}^s \vert \theta)$ the state-vector $x_{k+1}$ is replaced by $\widehat{x}_{k+1\vert n}^s$, the matrix identity (\ref{eq:eq14}) can not be used to simplify (\ref{eq:eq22}) to $\mathcal{I}_{\xi}^{-1}(\theta)$.  
\end{remark}

\subsubsection{Newton-Raphson iterative estimator}
In general, we can not directly employ (\ref{eq:eq21}) to calculate $\widehat{x}_{k\vert n}^s$ since it involves the true state-vector $x_k$ and the expected information matrix $\mathcal{I}_{\xi}(\theta)$. The section below discusses iterative schemes for ML smoothing. Replacing $x_k$ in (\ref{eq:eq21}) by $\widehat{x}_{k\vert n}^{s(\ell)}$, an estimate of $x_k$ obtained after $\ell-$iteration, $\ell\geq 1$, and $\widehat{x}_{k\vert n}^s$ by $\widehat{x}_{k\vert n}^{s(\ell +1)}$ while $\mathcal{I}_{\xi}(\theta)$ replaced by $J_{x_k,x_k}^{\xi}(\widehat{x}_k^{s(\ell)},\widehat{x}_{k+1\vert n}^s,y_{0:n}\vert\theta)$ yields
\begin{equation}\label{eq:algo1}
\begin{split}
\widehat{x}_{k\vert n}^{s(\ell +1)}=& \widehat{x}_{k\vert n}^{s(\ell)} + [J_{x_k,x_k}^{\xi}(\widehat{x}_{k\vert n}^{s(\ell)},\widehat{x}_{k+1\vert n}^s, y_{0:n} \vert \theta)]^{-1} \\
&\hspace{1cm}\times \mathcal{S}_{x_k}(\widehat{x}_{k\vert n}^{s(\ell)},\widehat{x}_{k+1\vert n}^s\vert \theta).
\end{split}
\end{equation}
This iteration is run until a stopping criterion is met,
%%
%\begin{align}\label{eq:stoppingrule}
%\frac{\Vert \widehat{x}_{k\vert n}^{s(\ell +1)} -  \widehat{x}_{k\vert n}^{s(\ell)} \Vert}{\Vert \widehat{x}_{k\vert n}^{s(\ell)} \Vert}  <\epsilon,
%\end{align}
%%
%
\begin{align}\label{eq:stoppingrule}
\Vert \widehat{x}_{k\vert n}^{s(\ell +1)} -  \widehat{x}_{k\vert n}^{s(\ell)} \Vert  <\epsilon,
\end{align}
for $\epsilon>0$. At convergence, the limiting smoother $\widehat{x}_{k\vert n}^{s(\infty)}$ corresponds to ML estimate as $ \mathcal{S}(\widehat{x}_{k\vert n}^{s(\infty)},\widehat{x}_{k+1\vert n}^s\vert \theta)=0$. 
To run iteration (\ref{eq:algo1}), we choose an initial condition:
\begin{align}\label{eq:initialcond}
\widehat{x}_{k\vert n}^{s(0)}=\mathbb{E}[x_k\vert \underline{y}_k,\theta].
\end{align}
This condition can be set up using particle filter, see Section \ref{sec:MCMC}. It would result in $\widehat{x}_{k\vert n}^s$ being a smoothed minimum variance unbiased estimate of state-vector $x_k$. 

\subsubsection{EM-gradient iterative estimator}

By majorant property (\ref{eq:eq10}) of $J_{x_k,x_k}^z(\widehat{x}_{k\vert n}^{s(\ell)},\widehat{x}_{k+1\vert n}^s, y_{0:n}\vert \theta)$ over   $J_{x_k,x_k}^{\xi}(\widehat{x}_{k\vert n}^{s(\ell)},\widehat{x}_{k+1\vert n}^s, y_{0:n}\vert \theta)$, the convergence of (\ref{eq:algo1}) may be increased by replacing $J_{x_k,x_k}^{\xi}(\widehat{x}_{k\vert n}^{s(\ell)},\widehat{x}_{k+1\vert n}^s, y_{0:n}\vert \theta)>0$ with $J_{x_k,x_k}^z(\widehat{x}_{k\vert n}^{s(\ell)},\widehat{x}_{k+1\vert n}^s, y_{0:n}\vert \theta)$ to get a recursion
\begin{equation}\label{eq:algo2}
\begin{split}
\widehat{x}_{k\vert n}^{s(\ell +1)}=& \widehat{x}_{k\vert n}^{s(\ell)} +[J_{x_k,x_k}^{z}(\widehat{x}_{k\vert n}^{s(\ell)},\widehat{x}_{k+1\vert n}^s, y_{0:n}\vert \theta)]^{-1}\\
 &\hspace{1cm}\times \mathcal{S}_{x_k}(\widehat{x}_{k\vert n}^{s(\ell)},\widehat{x}_{k+1\vert n}^s\vert \theta),
 \end{split}
\end{equation}
with faster convergence of $\{\widehat{x}_{k\vert n}^{s(\ell)}\}_{\ell \geq 0}$ than that of (\ref{eq:algo1}). 
%The update $\widehat{x}_k^{(\ell +1)}$ (\ref{eq:algo2}) corresponds to the maximizer of quadratic function $\mathcal{Q}\big(x_k\vert \widehat{x}_k^{(\ell)}\big)$ defined by 
%%
%\begin{align*}
%\mathcal{Q}(x_k\vert \widehat{x}_k^{(\ell)})=&\mathcal{Q}(\widehat{x}_k^{(\ell)}\vert \widehat{x}_k^{(\ell)}) + \mathcal{S}(\widehat{x}_k^{(\ell)},\underline{y}_k\vert\theta)\big(x_k-\widehat{x}_k^{(\ell)}\big)\\
%&-\frac{1}{2}\big(x_k-\widehat{x}_k^{(\ell)}\big)^{\top} J_z(\widehat{x}_k^{(\ell)},\underline{y}_k\vert\theta)\big(x_k-\widehat{x}_k^{(\ell)}\big),
%\end{align*}
%%
%such that the update $\widehat{x}_k^{(\ell +1)}$ (\ref{eq:algo2}) is read as
%\begin{align}\label{eq:EMfunc}
%\widehat{x}_k^{(\ell+1)}=\argmax_{x_k} \mathcal{Q}(x_k\vert \widehat{x}_k^{(\ell)}).
%\end{align}
%%

The recursive equation (\ref{eq:algo2}) is run on the same initial condition (\ref{eq:initialcond}). When the stopping criterion (\ref{eq:stoppingrule}) is reached at $\widehat{x}_{k\vert n}^{s(\infty)}$, $\mathcal{S}_{x_k}(\widehat{x}_{k\vert n}^{s(\infty)},\widehat{x}_{k+1\vert n}^s\vert \theta)=0$. Thus, $\widehat{x}_{k\vert n}^{s(\infty)}$ corresponds to the ML smoother $\widehat{x}_{k\vert n}^s$ (\ref{eq:maximizer2}). It extends further the EM-gradient algorithm of \cite{Lange} for ML parameter estimation to ML smoothing estimation.

\begin{remark}
To deal with invertibility of $J_{x_k,x_k}^{z}(\xi\vert \theta)$ in (\ref{eq:algo2}), we may consider replacing $J_{x_k,x_k}^{z}(\xi\vert \theta)$ by matrix
\begin{align*}
M_z(\xi\vert \theta)\equiv \mathbb{E}\big[\frac{\partial \log f(z\vert \theta)}{\partial x_k} \frac{\partial \log f(z \vert \theta)}{\partial x_k^{\top}}    \big\vert \xi,\theta \big].
\end{align*}
Note that by (\ref{eq:eq14}) both $J_{x_k,x_k}^z(\xi\vert \theta)$ and $M_z(\xi \vert \theta)$ have equal expectation. Replacing $J_{x_k,x_k}^z(\xi\vert \theta)$ by $M_z(\xi \vert \theta)$ in (\ref{eq:algo2}),
\begin{equation}\label{eq:algo3}
\begin{split}
\widehat{x}_{k\vert n}^{s(\ell +1)}=& \widehat{x}_{k\vert n}^{s(\ell)} +M_z^{-1}(\widehat{x}_{k\vert n}^{s(\ell)},\widehat{x}_{k+1\vert n}^s, y_{0:n}\vert \theta) \\
&\hspace{1cm}\times \mathcal{S}(\widehat{x}_{k\vert n}^{s(\ell)},\widehat{x}_{k+1\vert n}^s, y_{0:n}\vert \theta).
\end{split}
\end{equation}
At convergence, $\widehat{x}_{k\vert n}^{s(\infty)}$ corresponds to the ML smoother (\ref{eq:eq20}) since it satisfies $\mathcal{S}(\widehat{x}_{k\vert n}^{s(\infty)},\widehat{x}_{k+1\vert n}^s\vert \theta)=0$. Thus, iteration (\ref{eq:algo3}) extends further the BHHH algorithm of \cite{BHHH} and EM-gradient algorithm of \cite{Lange} for recursive ML smoothing estimation of state-vector $x_k$.
\end{remark}

\subsubsection{Locally equivalent EM algorithm}

The result below shows that the recursive equation (\ref{eq:algo2}) is locally equivalent to an EM algorithm for smoothing, which in turn extends the EM-algorithm of \cite{Dempster} for ML smoothing state estimation.
\begin{theorem}[\textbf{EM-algorithm}]
The recursive equation (\ref{eq:algo2}) is locally equivalent to the EM algorithm,
\begin{eqnarray*}
\widehat{x}_{k\vert n}^{s(\ell+1)}=\argmax_{x_k} \mathbb{E}\big[\log f(x_{0:n}, y_{0:n}\vert \theta) \big\vert \widehat{x}_{k\vert n}^{s(\ell)}, \widehat{x}_{k+1\vert n}^s,y_{0:n}, \theta\big].
\end{eqnarray*}
\end{theorem}
\vspace{-0.25cm}
The notion of being locally equivalent between (\ref{eq:algo2}) and the EM algorithm refers to the former having quadratic convergence compared to linear convergence of the latter. See Section 3.9 of \cite{McLachlan} and \cite{Lange} for details (on parameter estimation). 

Observe that the above EM-algorithm is slightly different from that of \cite{Ramadan}. 

\noindent \textbf{Proof.}
Since $\widehat{x}_{k\vert n}^{s(\ell +1)}$ is the maximizer of EM criterion, the proof follows from applying first-order Taylor expansion around current estimate $\widehat{x}_{k\vert n}^{s(\ell)}$ to work out
\begin{align*}
&\hspace{-0.5cm}\frac{\partial \log f(\widehat{x}_{k\vert n}^{s(\ell+1)},x_{k+1},\eta,y_{0:n}\vert \theta)}{\partial x_k}=\frac{\partial \log f(\widehat{x}_{k\vert n}^{s(\ell)},x_{k+1}, \eta, y_{0:n}\vert \theta)}{\partial x_k} \\
&\hspace{0.5cm}+ \frac{\partial^2 \log f(\widehat{x}_{k\vert n}^{s(\ell)},x_{k+1},\eta, y_{0:n}\vert \theta)}{\partial x_k \partial x_k^{\top}}(\widehat{x}_{k\vert n}^{s(\ell+1)} - \widehat{x}_{k\vert n} ^{s(\ell)}).
\end{align*}
Taking conditional expectation $\mathbb{E}[\bullet \vert \widehat{x}_{k\vert n}^{s(\ell)},x_{k+1},y_{0:n}\theta]$ on both sides, the proof is complete on account that $\mathbb{E}\big[\frac{\partial \log f(\widehat{x}_{k\vert n}^{s(\ell+1)},x_{k+1},\eta,y_{0:n}\vert \theta)}{\partial x_k}\big\vert \widehat{x}_{k\vert n}^{s(\ell)},x_{k+1},y_{0:n},\theta\big]$$=0$ since $\widehat{x}_{k\vert n}^{s(\ell+1)}$ is the EM criterion maximizer and by identity (\ref{eq:mainidentity}) $\mathbb{E}\big[\frac{\partial \log f(\widehat{x}_{k\vert n}^{s(\ell)}, x_{k+1},\eta,y_{0:n}\vert \theta)}{\partial x_k}\big\vert \widehat{x}_{k\vert n}^{s(\ell)},x_{k+1},y_{0:n},\theta\big]=\frac{\partial \log f(\widehat{x}_{k\vert n}^{s(\ell)},x_{k+1},y_{0:n}\vert \theta)}{\partial x_k}$ followed by replacing $x_{k+1}$ by $\widehat{x}_{k+1\vert n}^s$ and rearranging the term after multiplying by the inverse matrix $\Big(-\frac{\partial^2 \log f(\widehat{x}_{k\vert n}^{s(\ell)},x_{k+1},\eta, y_{0:n}\vert \theta)}{\partial x_k \partial x_k^{\top}}\Big)^{-1}$. $\square$

The result below shows that the EM-algorithm increases the loglikelihood function of incomplete data.

\begin{proposition}
EM-algorithm increases the incomplete-data loglikelihood function $\log f(x_k,x_{k+1},y_{0:n}\vert \theta)$, i.e.,
\begin{align*}
\log f(\widehat{x}_{k\vert n}^{s(\ell+1)},\widehat{x}_{k+1\vert n}^s,y_{0:n}\vert \theta) > \log  f(\widehat{x}_{k\vert n}^{s(\ell)},\widehat{x}_{k+1\vert n}^s,y_{0:n}\vert \theta).
\end{align*}
\end{proposition}
\noindent \textbf{Proof.}
The proof is central to showing that the function $x_k\rightarrow \mathbb{E}\big[\log f(x_k,x_{k+1},\eta,y_{0:n}\vert \theta) \big\vert \widehat{x}_{k\vert n}^{s(\ell)},x_{k+1},y_{0:n},\theta\big]$
is decreasing in $x_k$, which can be established using Jensen's inequality and on account that the likelihood ratio $\frac{f(\widehat{x}_{k\vert n}^{s(\ell +1)},x_{k+1},\eta,y_{0:n}\vert \widehat{x}_{k\vert n}^{s(\ell+1)},x_{k+1},y_{0:n},\theta)}{ f(\widehat{x}_{k\vert n}^{s(\ell)},x_{k+1},\eta,y_{0:n} \vert \widehat{x}_{k\vert n}^{s(\ell)},x_{k+1},y_{0:n},\theta ) } $ corresponds to the Radon-Nikodym derivative of changing underlying probability distribution with density function $ f(\widehat{x}_{k\vert n}^{s(\ell)},x_{k+1},\eta, y_{0:n}\vert \widehat{x}_{k\vert n}^{s(\ell)},x_{k+1},y_{0:n},\theta)$ to that of with density $ f(\widehat{x}_{k\vert n}^{s(\ell+1)},x_{k+1},\eta, y_{0:n}\vert \widehat{x}_{k\vert n}^{s(\ell+1)},x_{k+1},y_{0:n},\theta)$. $\square$

%On recalling that the ML smoother $\widehat{x}_{k\vert n}^s$ and its (estimated) covariance matrix are presented in terms of the score functions and observed information matrices, it remains to discuss valuation of the two quantities.

\subsection{Efficiency of the ML smoother $\widehat{x}_{k\vert n}^s$}

The results below show that as the solution of (\ref{eq:eq20}), the smoother $\widehat{x}_{k\vert n}^s$ corresponds to the value of $x_k$ that gives the smallest distance between $\log f(x_k,y_{0:k}\vert \theta_k)$ and $\log f(x_k^0,y_{0:k}\vert\theta_k)$, whereas $x_k^0$ is the true value of $x_k$.
\begin{proposition}\label{prop:efficiency}
As the solution of (\ref{eq:eq20}), it coincides with
\begin{align*}
\widehat{x}_{k\vert n}^s=\argmin_{x_k} \big\vert \log f(x_k,y_{0:k} \vert\theta) -  \log f(x_k^0,y_{0:k}\vert\theta) \big\vert^2.
\end{align*}
Furthermore, with $\widehat{x}_k=\argmax_{x_k} \log f(x_k,y_{0:k}\vert\theta)$, 
\begin{align*}
&\mathbb{E}\big[\big(\widehat{x}_{k\vert n}^s-x_k\big)\big(\widehat{x}_{k\vert n}^s-x_k\big)^{\top}\vert\theta\big] \\
&\hspace{1.5cm} < \mathbb{E}\big[\big(\widehat{x}_{k}-x_k\big)\big(\widehat{x}_{k}-x_k\big)^{\top}\vert\theta\big].
\end{align*}
\end{proposition}
\begin{remark} 
The above results show that $\widehat{x}_{k\vert n}^s$ (\ref{eq:maximizer2}) gives an estimate of $x_k^0$ with more adherence of loglikelihood with covariance matrix less than that of ML estimator $\widehat{x}_k$. 
%In the case of Gaussian linear system, see Section \ref{sec:sec4}, $\widehat{x}_{k\vert n}^s$ coincides with the MLE $\widehat{x}_k$ with less variance.
\end{remark} 

\noindent \textbf{Proof.}
See Appendix C for details of derivation. $\square$

\subsection{Sequential Monte Carlo method for valuation of score functions and observed information matrices}\label{sec:MCMC}

To this end, consider a sequence of random vectors $(x_{0:n},y_{0:n})$ satisfying the Markov assumption
\begin{equation}\label{eq:eq27}
\begin{split}
y_k\vert x_k \perp& (x_{0:k-1},y_{0:k-1}),\\
x_k\vert x_{k-1} \perp & (x_{0:k-1},y_{0:k-1}).
\end{split}
\end{equation}
By Markov property, the complete-data loglikelihood function of the random vectors $(x_{0:n},y_{0:n})$ is given by
\begin{align*}
&\log f(x_{0:n},y_{0:n}\vert \theta)=\log f(x_0\vert \theta)\\
&\hspace{1cm} + \sum_{k=0}^n \log f(y_k\vert x_k,\theta) + \sum_{k=1}^n\log f(x_k\vert x_{k-1},\theta).
\end{align*}
The \textit{complete-data score function} is determined by 
\begin{equation}\label{eq:eq28}
\begin{split}
&\frac{\partial \log f(x_{0:n},y_{0:n}\vert \theta)}{\partial x_k}= \frac{\partial \log f(y_k\vert x_k,\theta)}{\partial x_k}\\
&\hspace{1cm}+\frac{\partial \log f(x_{k+1}\vert x_k,\theta)}{\partial x_k} + \frac{\partial f(x_k\vert x_{k-1},\theta)}{\partial x_k}
\end{split}
\end{equation}
for $1\leq k\leq n-1$. In particle filtering literature, see e.g. \cite{Kitagawa,Kitagawa2021}, \cite{Godsill} and \cite{Doucet}, filtering distribution $f(x_k\vert y_{0:k},\theta)$ is presented as
\begin{align}\label{eq:eq29}
f(x_k\vert y_{0:k},\theta)=\sum_{m=1}^M \alpha_k^m \delta_{x_k^m}(x_k),
\end{align}
where $M>1$, $\delta_y(x)$ is the Dirac delta function , whilst 
\begin{align*}
\alpha_k^m=\frac{f(y_k\vert x_k^m,\theta)}{\sum_{m=1}^M f(y_k\vert x_k^m,\theta)}
\end{align*}
denotes the probability of drawing a sample $x_k^m$ from $f(x_k\vert y_{0:k},\theta)$. Using (\ref{eq:eq29}), a minimum square error state estimator $\widehat{x}_{k\vert k}=\mathbb{E}[x_k\vert y_{0:k},\theta]$ is given by $\sum_{m=1}^m \alpha_k^m x_k^m$.

For the valuation of score function and observed information matrices, the result below is required.
\begin{proposition}
Consider probability distribution 
\begin{align*}
w_{k-1}^m(x_k\vert \theta)=\frac{\alpha_{k-1}^m f(x_k\vert x_{k-1}^m,\theta)}{\sum_{m=1}^M \alpha_{k-1}^m f(x_k\vert x_{k-1}^m,\theta)}.
\end{align*}
Under (\ref{eq:eq27}), distribution $f(x_{k-1}\vert x_k,x_{k+1},y_{0:n},\theta)$ reads 
%%
%\begin{equation}\label{eq:eq30}
%\begin{split}
%&f(x_{k-1}\vert x_k,x_{k+1},y_{0:n},\theta)=f(x_{k-1}\vert x_k,y_{0:k-1},\theta)\\
%&\hspace{1cm}=\sum_{m=1}^M w_{k-1}^m(x_k\vert \theta) \delta_{x_{k-1}^m}(x_{k-1}).
%\end{split}
%\end{equation}
%%%
\begin{equation*}
\begin{split}
f(x_{k-1}\vert x_k,x_{k+1},y_{0:n},\theta)=\sum_{m=1}^M w_{k-1}^m(x_k\vert \theta) \delta_{x_{k-1}^m}(x_{k-1}).
\end{split}
\end{equation*}

\end{proposition}
\noindent \textbf{Proof.} By the Markov property (\ref{eq:eq27}), Bayes formula and (\ref{eq:eq27}), one can show that $f(y_{k:n}\vert x_k,x_{k-1},y_{0:k-1},\theta)=f(y_{k:n}\vert x_k,y_{0:k-1},\theta)$ resulting in $f(x_{k-1}\vert x_k,x_{k+1},y_{0:n},\theta)\linebreak=f(x_{k-1}\vert x_k,y_{0:k-1},\theta).$ The proof is complete on account of $f(x_{k-1}\vert x_k,y_{0:k-1},\theta)=\frac{f(x_{k-1}\vert y_{0:k-1},\theta)f(x_k\vert x_{k-1},\theta)}{f(x_k\vert y_{0:k-1},\theta)}$,
%%
%\begin{align*}
%&f(x_{k-1}\vert x_k,y_{0:n},\theta)=f(x_{k-1}\vert x_k,y_{0:k-1},\theta)\\
%&=\frac{f(x_{k-1}\vert y_{0:k-1},\theta)f(x_k\vert x_{k-1},\theta)}{f(x_k\vert y_{0:k-1},\theta)},
%\end{align*}
%%
%$1\leq k\leq n$. Let $\mathcal{X}_{k-1}$ be the set of any possible values of $x_{k-1}$. By the law of total probability and Bayes formula,
%
\begin{align*}
f(x_k\vert y_{0:k-1},\theta)=&\int_{\mathcal{X}_{k-1}} f(x_k\vert x_{k-1},\theta)f(x_{k-1}\vert y_{0:k-1},\theta) d x_{k-1}\\
%&=\sum_{m=1}^M\ \alpha_{k-1}^m \int_{\mathcal{X}_{k-1}} f(x_k\vert x_{k-1},y_{0:k-1},\theta) \delta_{x_{k-1}^m}(x_{k-1}) d x_{k-1}\\
=&\sum_{m=1}^M \alpha_{k-1}^m f(x_k\vert x_{k-1}^m,\theta). \quad \square
\end{align*}
%
%which in turn establishes the claim. $\square$
Notice that in contrary to the smoothing distribution $f(x_k\vert y_{0:n},\theta)$ discussed in \cite{Hurzeler}, \cite{Doucet2000}, \cite{Godsill2004}, \cite{Vo} and \cite{Saha}, $f(x_{k-1}\vert x_k,x_{k+1},y_{0:n},\theta)$ does not involve importance weight $\alpha_{k\vert n}^m$ in the scheme\footnote{defined by $\alpha_{k\vert n}^j=\sum_{m=1}^M \alpha_{k+1\vert n}^m \frac{\alpha_k^j f(x_{k+1}^m\vert x_k^j,\theta)}{\sum_{\ell=1}^M \alpha_k^{\ell} f(x_{k+1}^m\vert x_k^{\ell},\theta)}$, with $\alpha_{n\vert n}^m =\alpha_n^m$. It is used to specify smoothing distribution $f(x_k\vert y_{0:n},\theta)=\sum_{m=1}^M \alpha_{k\vert n}^m \delta_{x_k^m}(x_k)$. See \cite{Doucet2000}. This smoothing distribution was used in \cite{Saha} to find the ML smoother $\widehat{x}_{k\vert n}^s=\argmax_{x_k} f(x_k\vert y_{0:n},\theta)$.}.

\subsubsection{SMC valuation of the score function}
From (\ref{eq:eq28}) the incomplete-data score function reads
\begin{eqnarray}\label{eq:eq31}
&\mathcal{S}_{x_k}(x_k,x_{k+1}\vert\theta)=\mathbb{E}\big[\frac{\partial \log f(x_{0:n},y_{0:n}\vert \theta)}{\partial x_k} \big\vert x_k,x_{k+1},y_{0:n},\theta\big] \nonumber\\
&=\frac{\partial \log f(y_k\vert x_k,\theta)}{\partial x_k} + \frac{\partial \log f(x_{k+1}\vert x_k,\theta)}{\partial x_k}\nonumber\\
&+\mathbb{E}\big[\frac{\partial \log f(x_k\vert x_{k-1},\theta)}{\partial x_k}\big\vert x_k,x_{k+1},y_{0:n},\theta\big].
\end{eqnarray}
The last conditional expectation is evaluated as
\begin{align}\label{eq:condexp}
&\mathbb{E}\big[\frac{\partial \log f(x_k\vert x_{k-1},\theta)}{\partial x_k} \big\vert x_k,x_{k+1},y_{0:n},\theta\big]  \nonumber\\
&=\int_{\mathcal{X}_{k-1}} \frac{\partial \log f(x_k\vert x_{k-1},\theta)}{\partial x_k}  f(x_{k-1}\vert x_k,x_{k+1},y_{0:n},\theta) d x_{k-1} \nonumber\\
&=\sum_{m=1}^M w_{k-1}^m(x_k\vert \theta) \frac{\partial \log f(x_k\vert x_{k-1}^m,\theta)}{\partial x_k}.
\end{align}

\subsubsection{SMC valuation of observed information matrices}
The observed information matrix $J_{x_k,x_k}^z(x_k,x_{k+1},y_{0:n}\vert\theta)$ used in iterative scheme (\ref{eq:algo2}) is evaluated from (\ref{eq:eq28}) as
\begin{align*}
&J_{x_k,x_k}^z(x_k,x_{k+1},y_{0:n}\vert\theta)\\
=&\mathbb{E}\big[-\frac{\partial^2 \log f(x_{0:n},y_{0:n}\vert\theta)}{\partial x_k \partial x_k^{\top}}\big\vert x_k,x_{k+1},y_{0:n},\theta\big]\\
=&-\frac{\partial^2 \log f(y_k\vert x_k,\theta)}{\partial x_k\partial x_k^{\top}}-\frac{\partial^2 \log f(x_{k+1}\vert x_k,\theta)}{\partial x_k \partial x_k^{\top}}\\
&+\mathbb{E}\big[-\frac{\partial^2 \log f(x_k\vert x_{k-1},\theta)}{\partial x_k \partial x_k^{\top}}\big\vert x_k,x_{k+1},y_{0:n},\theta\big].
\end{align*}
%where the last conditional expectation is evaluated as
%%
%\begin{align*}
%&\mathbb{E}\big[-\frac{\partial^2 \log f(x_k\vert x_{k-1},\theta)}{\partial x_k \partial x_k^{\top}}\big\vert x_k,x_{k+1},y_{0:n},\theta\big]\\
%&=-\sum_{m=1}^M w_{k-1}^m(x_k\vert\theta) \frac{\partial^2 \log f(x_k\vert x_{k-1}^m,\theta)}{\partial x_k \partial x_k^{\top}}.
%\end{align*}
%%
The observed information matrix $J_{x_k,x_k}^{\xi}(x_k,x_{k+1},y_{0:n}\vert\theta)$ is evaluated either using identity (\ref{eq:louis1}) or by directly taking derivative w.r.t $x_k$ of (\ref{eq:eq31}). Either way, the information matrix $J_{x_k,x_k}^{\xi}(x_k,x_{k+1},y_{0:n}\vert\theta)$ is given by 
\begin{eqnarray}\label{eq:eq32}
&J_{x_k,x_k}^{\xi}(x_k,x_{k+1},y_{0:n}\vert\theta)=J_{x_k,x_k}^z(x_k,x_{k+1},y_{0:n}\vert\theta) \nonumber \\
&-\mathbb{E}\big[\frac{\partial \log f(x_k\vert x_{k-1},\theta)}{\partial x_k} \frac{\partial \log f(x_k\vert x_{k-1},\theta)}{\partial x_k^{\top}}\big\vert x_k,x_{k+1},y_{0:n},\theta\big] \nonumber\\
&+\Big(\mathbb{E}\big[\frac{\partial \log f(x_k\vert x_{k-1},\theta)}{\partial x_k}\big\vert x_k,x_{k+1},y_{0:n},\theta\big]\Big)\\
&\hspace{1cm}\times \Big(\mathbb{E}\big[\frac{\partial \log f(x_k\vert x_{k-1},\theta)}{\partial x_k^{\top}}\big\vert x_k,x_{k+1},y_{0:n},\theta\big]\Big). \nonumber
\end{eqnarray}
%
%The first conditional expectation is evaluated as
%%
%\begin{align*}
%&\mathbb{E}\big[\frac{\partial \log f(x_k\vert x_{k-1},\theta)}{\partial x_k} \frac{\partial \log f(x_k\vert x_{k-1},\theta)}{\partial x_k^{\top}}\big\vert x_k,x_{k+1},y_{0:n},\theta\big]\\
%&=\sum_{m=1}^M w_{k-1}^m(x_k\vert\theta)\frac{\partial \log f(x_k\vert x_{k-1}^m,\theta)}{\partial x_k} \frac{\partial \log f(x_k\vert x_{k-1}^m,\theta)}{\partial x_k^{\top}},
%\end{align*}
%%
All conditional expectations are evaluated using the distribution $f(x_{k-1}\vert x_k,x_{k+1},y_{0:n},\theta)$ similarly to (\ref{eq:condexp}). 

For the valuation of covariance matrix $\Sigma_{x_k,x_k}(\theta)$ and its estimate $\widehat{\Sigma}_{k\vert n}^s$, we need to evaluate the information matrix $J_{x_{k+1},x_k}^{\xi}(x_k,x_{k+1},y_{0:n}\vert\theta)$. From the complete-data score function (\ref{eq:eq28}) and (\ref{eq:eq31}), the matrix is given by
\begin{align*}
J_{x_{k+1},x_k}^{\xi}(x_k,x_{k+1},y_{0:n}\vert\theta)=-\frac{\partial^2 \log f(x_{k+1}\vert x_k,\theta)}{\partial x_{k+1}\partial x_k^{\top}}.
\end{align*}
The observed information matrix $J_{x_k,x_{k+1}}^{\xi}(x_k,x_{k+1},y_{0:n}\vert\theta)$ is derived from  (\ref{eq:eq28}). However, since by the Markov property (\ref{eq:eq27}) $f(x_{k+2}\vert x_k,x_{k+1},y_{0:n},\theta)=f(x_{k+2}\vert x_{k+1},y_{0:n},\theta)$,
\begin{align*}
J_{x_k,x_{k+1}}^{\xi}(x_k,x_{k+1},y_{0:n}\vert\theta)=-\frac{\partial^2 \log f(x_{k+1}\vert x_k,\theta)}{\partial x_k \partial x_{k+1}^{\top}}.
\end{align*}

\subsection{Repeated sampling method}\label{sec:mcmc2}
To derive an estimate of the covariance matrix of $\widehat{x}_{k\vert n}^{s(\infty)}-x_k$, repeated sampling is used to draw $x_{k-1}\sim f(x_{k-1}\vert x_k,\underline{y}_k,\theta)$ into $N$ random samples of size $M$ each. To each subsample $\{x_{k-1,\ell}^{m}: m=1,\ldots,M\}$, $\ell \in \{1,\ldots,N\}$, apply iteration (\ref{eq:algo2}) until convergence to  $\widehat{x}_{k,\ell\vert n}^{s}$. After $N$ samples having been used, we have $N$ estimators $\{\widehat{x}_{k,\ell\vert n}^{s}\}$, $\{J_{x_k,x_k}^{\xi(\ell)}(\xi_{k\vert n}^{(\ell)}\vert\theta)\}$ and $\{J_{x_k,x_{k+1}}^{\xi(\ell)}(\widehat{\xi}_{k\vert n}^{(\ell)}\vert\theta)\}$ with $\widehat{\xi}_{k\vert n}^{(\ell)}:=(\widehat{x}_{k,\ell\vert n}^{s}, \widehat{x}_{k+1,\ell\vert n}^s,y_{0:n})$. The smoothed estimates of $x_k$ and $I_{\xi}(\theta)$ are respectively 
\begin{align*}
\widehat{x}_k^0:=&\frac{1}{N}\sum_{\ell=1}^N \widehat{x}_{k,\ell\vert n}^{s}, \\
\widehat{I}_{x_k,x_k}^{\xi}(\theta):=&\frac{1}{N}\sum_{\ell=1}^N J_{x_k,x_{k}}^{\xi(\ell)}(\widehat{\xi}_{k\vert n}^{(\ell)}\vert\theta),
\end{align*}
with $I_{x_k,x_k}^{\xi}(\theta):=\mathbb{E}[J_{x_k,x_{k}}^{\xi}(\xi_k\vert\theta)\vert\theta]$. Similarly defined for $\widehat{I}_{x_k,x_{k+1}}^{\xi}(\theta)$.
Thus, an estimate of $\Sigma_{x_k,x_k}^s(\theta)$ is given by 
\begin{align}
\widehat{\Sigma}_{x_k,x_k}^s(\theta)=&[\widehat{I}_{x_k,x_k}^{\xi}(\theta)]^{-1}\widehat{I}_{x_k,x_{k+1}}^{\xi}(\theta) \widehat{\Sigma}_{x_{k+1},x_{k+1}}^s(\theta) \nonumber \\
&\hspace{-1cm}\times \widehat{I}_{x_{k+1},x_k}^{\xi}(\theta)[\widehat{I}_{x_k,x_k}^{\xi}(\theta)]^{-1}+[\widehat{I}_{x_k,x_k}^{\xi}(\theta)]^{-1} \label{eq:estvar}.
\end{align}

\section{Application of the main results}\label{sec:sec4}

\subsection{Linear state-space models}\label{sec:application}
Suppose that a controlled state-vector $x_k\in \mathbb{R}^p$, $p\geq1$, evolves according to a recursive linear system
\begin{align}\label{eq:eq35}
x_k=F_k x_{k-1} + G_k u_k + v_k,
\end{align}
where $u_k\in\mathbb{R}^m$, $m\leq p$, forms a sequence of predetermined control variable. The state-vector $x_k$ is assumed to be observed through noisy measurement $y_k\in\mathbb{R}^q$,
\begin{align}\label{eq:eq36}
y_k=H_k x_k + w_k,
\end{align}
where $q\leq p$, $v_k$ and $w_k$ are independent sequence of random vectors with $v_k\sim N(0,Q_k)$ and $w_k\sim N(0,R_k)$. Assume that all matrices $F_k$, $G_k$, $H_k$, $Q_k$ and $R_k$ are conformable and prespecified. Also, the initial state $x_0\sim N(\mu,P_0)$. Let $\theta=(F_{\ell},G_{\ell},H_{\ell}, Q_{\ell}, R_{\ell}:\ell=1,\ldots,n)\cup (\mu,P_0)$ denote parameters of the system (\ref{eq:eq35})-(\ref{eq:eq36}).

\subsubsection{Loglikelihood and score functions}

From the linear system (\ref{eq:eq35})-(\ref{eq:eq36}) we deduce that
\begin{align*}
\log f(y_k\vert x_k,\theta)=&-\frac{1}{2}\big(\Vert y_k-H_k x_k\Vert_{R_k^{-1}}^2 +\log(2\pi \vert R_k\vert)\big),\\
\log f(x_k\vert x_{k-1},\theta)=&-\frac{1}{2}\big(\Vert x_k-F_k x_{k-1} -G_k u_k\Vert_{Q_k^{-1}}^2\\
&+ \log(2\pi \vert Q_k\vert)\big),
\end{align*}
with $\Vert x\Vert_A^2:=x^{\top}Ax$ and complete-data score function
\begin{align}\label{eq:scoreLN}
&\frac{\partial \log f(x_{0:n},y_{0:n}\vert\theta)}{\partial x_k} =H_k^{\top}R_k^{-1}(y_k-H_k x_k)  \nonumber \\
&\hspace{1cm}+F_{k+1}^{\top}Q_{k+1}^{-1}(x_{k+1}-F_{k+1}x_k -G_{k+1}u_{k+1})  \nonumber\\
&\hspace{1cm}-Q_k^{-1}(x_k-F_k x_{k-1}-G_k u_k).
\end{align}
The incomplete-data score function is specified by
\begin{align*}
&\mathcal{S}_{x_k}(x_k,x_{k+1}\vert\theta)=H_k^{\top}R_k^{-1}(y_k-H_k x_k)\\
&\hspace{1cm}+F_{k+1}^{\top}Q_{k+1}^{-1}(x_{k+1}-F_{k+1}x_k -G_{k+1}u_{k+1})\\
&\hspace{1cm}-Q_k^{-1}(x_k-F_k\mathbb{E}[x_{k-1}\vert x_k,x_{k+1},y_{0:n},\theta]-G_k u_k).
\end{align*}
The conditional expectation of $x_{k-1}$ can be worked out using the the distribution function $f(x_{k-1}\vert x_k,x_{k+1},y_{0:n},\theta)$,
\begin{align*}
&\mathbb{E}[x_{k-1}\vert x_k,x_{k+1}, y_{0:n},\theta]=\mathbb{E}[x_{k-1}\vert x_k,y_{0:k-1},\theta]\\
&=\widehat{x}_{k-1\vert k-1} +\Sigma_{k\vert k-1} P_{k\vert k}^{-1}(x_k-\widehat{x}_{k\vert k-1}),
\end{align*}
with $\Sigma_{k\vert k-1}:=\mathbb{E}[(x_{k-1}-\widehat{x}_{k-1\vert k-1})(x_k-\widehat{x}_{k\vert k-1})^{\top}\vert \theta]$,
\begin{align*}
\Sigma_{k+1\vert k}=&P_{k\vert k-1}(F_{k+1} - F_{k+1}K_k H_k)^{\top},\\
F_k\Sigma_{k\vert k-1}=&P_{k\vert k-1}-Q_k,
\end{align*}
where $P_{k\vert k-1}$ is the covariance matrix of $(\widehat{x}_{k\vert k-1}-x_k)$. See \cite{Surya2022b} for details. This result is used to replace the distribution $f(x_k\vert y_{0:k},\theta)$ widely used in filtering literature explained in the Introduction. Denoting by $P_{k\vert k}$ the covariance matrix of $\widehat{x}_{k\vert k}-x_k$, an application of the Woodbury matrix inversion formula yields
\begin{align*}
H_k^{\top} R_k^{-1} H_k + P_{k\vert k-1}^{-1}=P_{k\vert k}^{-1}.
\end{align*}
see p.258 in \cite{Higham}, which leads to an explicit form of the incomplete-data score function given by
\begin{align}\label{eq:eq37}
\mathcal{S}_{x_k}(x_k,x_{k+1}\vert\theta)=&P_{k\vert k}^{-1} \widehat{x}_{k\vert k} - [P_{k\vert k}^{-1}+ F_{k+1}^{\top}Q_{k+1}^{-1}F_{k+1}]x_k \nonumber\\
&\hspace{-0.5cm}+ F_{k+1}^{\top}Q_{k+1}^{-1} \big(x_{k+1} - G_{k+1}u_{k+1}\big),
\end{align}
and subsequently the observed information matrix 
\begin{align*}
J_{x_k,x_k}^{\xi}(x_k,x_{k+1},y_{0:n}\vert \theta)=P_{k\vert k}^{-1}+ F_{k+1}^{\top}Q_{k+1}^{-1}F_{k+1}>0.
\end{align*}
It follows from (\ref{eq:eq37}) that $\mathbb{E}\big[\mathcal{S}_{x_k}(x_k,x_{k+1}\vert\theta)\vert x_k,y_{0:k},\theta\big]=P_{k\vert k}^{-1}(\widehat{x}_{k\vert k}-x_k)=\frac{\partial \log f(x_k,y_{0:k}\vert \theta)}{\partial x_k}$, with $f(x_k\vert y_{0:k},\theta)=N(x_k\vert \widehat{x}_{k\vert k}, P_{k\vert k}^{-1})$, the filtering distribution used in \cite{Rauch} to derive ML estimators $\widehat{x}_{k\vert k}$ and $\widehat{x}_{k\vert n}^s$.

\subsubsection{Derivation of RTS smoother}
By setting the score function $\mathcal{S}_{x_k}(x_k,\widehat{x}_{k+1\vert n}^s\vert\theta)$ to zero and taking account of the following matrix inversion, derived using Woodbury matrix inversion formula,
\begin{align*}
[P_{k\vert k}^{-1}+ F_{k+1}^{\top}Q_{k+1}^{-1}F_{k+1}]^{-1}=&P_{k\vert k} - C_k F_{k+1}P_{k\vert k},\\
[P_{k\vert k} -C_k F_{k+1}P_{k\vert k}]F_{k+1}^{\top} Q_{k+1}^{-1}=&C_k,
\end{align*}
with $C_k=P_{k\vert k}F_{k+1}^{\top}P_{k+1\vert k}^{-1}$ and $P_{k+1\vert k}=F_{k+1}P_{k\vert k}F_{k+1}^{\top} + Q_{k+1}$, the ML smoother $\widehat{x}_{k\vert n}^s$ is finally given by
\begin{align}\label{eq:eq38}
\widehat{x}_{k\vert n}^s= \widehat{x}_{k\vert k} + C_k (\widehat{x}_{k+1\vert n}^s - \widehat{x}_{k+1\vert k}),
\end{align}
where $\widehat{x}_{k+1\vert k}=F_{k+1} \widehat{x}_{k\vert k} + G_{k+1}u_{k+1}$ defines the a priori estimate of $x_k$. The smoother $\widehat{x}_{k\vert n}^s$ takes the same form as the  \cite{Rauch} RTS smoother. The same smoothing solution (\ref{eq:eq38}) can also be obtained from (\ref{eq:eq21}).

\subsubsection{Covariance matrix and efficiency of $\widehat{x}_{k\vert n}^s$}
To derive the covariance matrix $\Sigma_{k\vert n}^s=\mathbb{E}[(\widehat{x}_{k\vert n}^s -x_k)(\widehat{x}_{k\vert n}^s -x_k)^{\top}\vert \theta]$, recall following score function (\ref{eq:eq37}),
\begin{align*}
J_{x_k,x_{k+1}}^{\xi}(x_k,x_{k+1},y_{0:n}\vert\theta)=& F_{k+1}^{\top} Q_{k+1}^{-1},\\
J_{x_{k+1},x_k}^{\xi}(x_k,x_{k+1},y_{0:n}\vert\theta)=& Q_{k+1}^{-1}F_{k+1}.
\end{align*}
The covariance matrix formula (\ref{eq:eq16}) reads as
\begin{align}
&\Sigma_{k\vert n}^s=[\mathbf{I}^{\xi}]_{x_k,x_k}^{-1}(\theta)=(P_{k\vert k}-C_kF_{k+1}P_{k\vert k}) \nonumber\\
&\hspace{1cm}+(P_{k\vert k}-C_k F_{k+1}P_{k\vert k})F_{k+1}^{\top}Q_{k+1}^{-1} \nonumber \\
&\hspace{1cm}\times \Sigma_{k+1\vert n}^s[(P_{k\vert k}-C_k F_{k+1}P_{k\vert k})F_{k+1}^{\top} Q_{k+1}^{-1}]^{\top}  \nonumber \\
&\hspace{1cm}= P_{k\vert k} + C_k(\Sigma_{k+1\vert n}^s - P_{k+1\vert k})C_k^{\top}, \label{eq:rtsvar}
\end{align}
where the last equality is due to $C_k F_{k+1}P_{k\vert k}=C_k P_{k+1\vert k}C_k^{\top}$. This is exactly the covariance matrix of RTS smoother derived in \cite{Rauch} using (\ref{eq:eq38}). 

The same conclusion is reached from (\ref{eq:eq22}) taking account that $\mathcal{S}_{x_k}(x_k,\widehat{x}_{k+1\vert n}^s\vert\theta)=P_{k\vert k}^{-1}(\widehat{x}_{k\vert k}-x_k)+ F_{k+1}^{\top}Q_{k+1}^{-1}(\widehat{x}_{k+1\vert n}^s - x_{k+1}) + F_{k+1}^{\top} Q_{k+1}^{-1} v_{k+1}$ and mutually independence of $\widehat{x}_{k\vert k}-x_k$, $\widehat{x}_{k+1\vert n}^s -x_{k+1}$ and $v_{k+1}$. Hence, since the mean square error $\Sigma_{k\vert n}^s$ equals to the inverse of expected information matrix $[\mathbf{I}^{\xi}]_{x_k,x_k}^{-1}(\theta)$, the RTS smoother (\ref{eq:eq38}) is a fully efficient state estimator.

To verify the result of Proposition \ref{prop:efficiency}, first note that $P_{k\vert k}=(I-K_k H_k)P_{k\vert k-1}$, $k\geq 1$, with $K_k$ being the Kalman gain defined by $K_k=P_{k\vert k-1}H_k^{\top}(H_k P_{k\vert k-1} H_k^{\top} + R_k)^{-1}$. One can check that $0<K_kH_k<I$ and therefore $0<P_{k\vert k}<P_{k\vert k-1}$, i.e., the a posterior covariance $P_{k\vert k}$ is less than the a priori covariance $P_{k\vert k-1}$ for all $k\geq 1$. Since $\Sigma_{n\vert n}^s=P_{n\vert n}$, then for $k=n-1$ it follows from (\ref{eq:rtsvar}) that $\Sigma_{n-1\vert n}^s= P_{n-1\vert n-1} + C_{n-1} (\Sigma_{n\vert n}^s -P_{n\vert n-1})C_{n-1}^{\top}<P_{n-1\vert n-1}.$ For $k=n-2$, we have $\Sigma_{n-2\vert n}^s= P_{n-2\vert n-2} + C_{n-2}(\Sigma_{n-1\vert n}^s - P_{n-1\vert n-2})C_{n-2}^{\top}<P_{n-2\vert n-2} +C_{n-2}(P_{n-1\vert n-1} - P_{n-1\vert n-2})C_{n-2}^{\top}<P_{n-2\vert n-2}.$ Finally, using induction argument shows that $$0<\Sigma_{k\vert n}^s <P_{k\vert k}. \quad \square$$
Thus, additional factor $(\widehat{x}_{k+1\vert n}^s -\widehat{x}_{k+1\vert n})$ in (\ref{eq:eq38}) serves as smoothing correcting term for the Kalman filter $\widehat{x}_{k\vert k}$.

%The fact that $\widehat{x}_{k\vert n}^s$ coincides with the Kalman estimator $\widehat{x}_{k\vert k}$ is that 

\subsection{Nonlinear state-space models}
This section derives iterative scheme (\ref{eq:algo2}) for a nonlinear state-space model with linear measurement
\begin{align*}
x_k=&F_k(x_{k-1},u_k) + v_k,\\
y_k=& H_k x_k +w_k,
\end{align*}
with $v_k\sim N(0,Q_k)$ and $w_k\sim N(0,R_k)$. Following these nonlinear systems we deduce that
\begin{align*}
\log f(y_k\vert x_k,\theta)=&-\frac{1}{2}\big(\Vert y_k - H_k x_k\Vert_{R_k^{-1}}^2 + \log(2\pi \vert R_k\vert)\big),\\
\log f(x_k\vert x_{k-1},\theta)=&-\frac{1}{2}\big(\Vert x_k -F_k(x_{k-1},u_k)\Vert_{Q_k^{-1}}^2\\
&+ \log (2\pi \vert Q_k\vert)\big).
\end{align*}

\subsubsection{Score functions of incomplete-data}
The complete-data  score function (\ref{eq:eq28}) reads as
\begin{align}\label{eq:eq39}
&\frac{\partial \log f(x_{0:n},y_{o:n}\vert \theta)}{\partial x_k}=H_k^{\top} R_k^{-1}(y_k-H_k x_k) \nonumber\\
&+\frac{\partial F_{k+1}^{\top}(x_k,u_{k+1})}{\partial x_k}Q_{k+1}^{-1}\big(x_{k+1}-F_{k+1}(x_k,u_{k+1})\big) \nonumber\\
&-Q_k^{-1}\big(x_k-F_k(x_{k-1},u_k)\big).
\end{align}
Thus, the incomplete-data score function is given by
\begin{align}\label{eq:scorenl}
&\mathcal{S}_{x_k}(x_k,x_{k+1}\vert\theta)=H_k^{\top}R_k^{-1}(y_k-H_k x_k) \nonumber\\
&+\frac{\partial F_{k+1}^{\top}(x_k,u_{k+1})}{\partial x_k}Q_{k+1}^{-1}\big(x_{k+1}-F_{k+1}(x_k,u_{k+1})\big) \nonumber\\
&-Q_k^{-1}\Big(x_k-\sum_{m=1}^M w_{k-1}^m(x_k\vert\theta) F_k(x_{k-1}^m,u_k)\Big).
\end{align}
\begin{remark}
The score function above posts difficulties in implementation of the EM algorithm. Particularly in getting an explicit update $\widehat{x}_{k\vert n}^{s(\ell+1)}$ from the M-step of the algorithm. This is due to nonlinearity of $F_{k+1}(x_k,u_{k+1})$.
\end{remark}

For this reason, the smoothing estimation is resolved using the EM-gradient-particle iterative scheme (\ref{eq:algo2}). 

\subsubsection{Observed information matrices}
Information matrix $J_{x_k,x_k}^z(x_k,x_{k+1},y_{0:n}\vert\theta)$ reads
\begin{align}\label{eq:eq40}
&J_{x_k,x_k}^z(x_k,x_{k+1},y_{0:n}\vert\theta)=H_k^{\top}R_k^{-1}H_k + Q_k^{-1} \nonumber\\
&+\Big(\frac{\partial F_{k+1}^{\top}(x_k,u_{k+1})}{\partial x_k}\Big)Q_{k+1}^{-1}\Big(\frac{\partial F_{k+1}^{\top}(x_k,u_{k+1})}{\partial x_k}\Big)^{\top} \nonumber\\
&-\Big(\frac{\partial^2 F_{k+1}^{\top}(x_k,u_{k+1})}{\partial x_k \partial x_k^{\top}}\Big)\big(Q_{k+1}^{-1}v_{k+1}\otimes I_{p\times p}\big),
\end{align}
where $v_{k+1}=x_{k+1}-F_{k+1}(x_k,u_{k+1})$ and $I_{p\times p}$ is a $(p\times p)-$identity matrix, whilst $\otimes$ denotes the Kronecker product of matrices. See for e.g. \cite{Khang} for details on partial derivative of matrix valued function.

For the calculation of covariance matrix (\ref{eq:eq16}) we need 
\begin{align*}
J_{x_k,x_{k+1}}^{\xi}(x_k,x_{k+1},y_{0:n}\vert\theta)=\Big(\frac{\partial F_{k+1}^{\top}(x_k,u_{k+1})}{\partial x_k}\Big)Q_{k+1}^{-1},
\end{align*}
thus, $J_{x_{k+1},x_{k}}^{\xi}(x_k,x_{k+1},y_{0:n}\vert\theta)=Q_{k+1}^{-1}\big(\frac{\partial F_{k+1}^{\top}(x_k,u_{k+1})}{\partial x_k}\big)^{\top}$. It remains to evaluate $J_{x_k,x_k}^{\xi}(x_k,x_{k+1},y_{0:n}\vert\theta)$. By (\ref{eq:eq32}),
\begin{align*}
&J_{x_k,x_k}^{\xi}(x_k,x_{k+1},y_{0:n}\vert\theta)=J_{x_k,x_k}^z(x_k,x_{k+1},y_{0:n}\vert\theta)\\
&-Q_k^{-1}\Big(\sum_{m=1}^M w_{k-1}^m(x_k\vert\theta)\big(x_k-F_k(x_{k-1}^m,u_k)\big)\\
&\hspace{3.5cm}\times \big(x_k-F_k(x_{k-1}^m,u_k)\big)^{\top}\Big)Q_k^{-1}\\
&+Q_k^{-1}\Big(\sum_{m=1}^M w_{k-1}^m(x_k\vert\theta)\big(x_k-F_k(x_{k-1}^m,u_k)\big)\Big)\\
&\hspace{1cm}\times \Big(\sum_{m=1}^M w_{k-1}^m(x_k\vert\theta)\big(x_k-F_k(x_{k-1}^m,u_k)\big)^{\top}\Big)Q_k^{-1}.
\end{align*}

\section{Numerical examples}\label{sec:numeric}

\subsection{Linear state-space model}

Consider the linear state-space (1) with the following parameter values, used in \cite{Ramadan}:
\begin{align*}
x_k=&\left(
\begin{array}{c}
x_{k,1}\\
x_{k,2}\\
x_{k,3}
\end{array}\right),
\;\;
F_k=\left(
\begin{array}{ccc}
0.66 & -1.31 & -1.11\\
0.07 & 0.73 & -0.06\\
0.00 & 0.08 & 0.80
\end{array}\right),\\
H_k=&\left(
\begin{array}{ccc}
0 & 1 & 1
\end{array}\right),
\;\;
Q_k=\left(
\begin{array}{ccc}
0.2 & 0 & 0\\
0  & 0.3 & 0\\
0 & 0 & 0.5
\end{array}\right),\\
R_k=&0.1, \;\;
\mu=\left(
\begin{array}{c}
0\\
0\\
0
\end{array}\right),
\;\;
P_0=\left(
\begin{array}{ccc}
0.3 & 0 & 0\\
0  & 0.3 & 0\\
0 & 0 & 0.3
\end{array}\right),
\end{align*}
in the absence of control variable $u_k$, i.e., $G_k=\mathbf{0}$. The state-vector $x_k$, started at a random initial $x_0\sim N(\mu,P_0)$, is partially observed through $y_k=H_k x_k + w_k$ with $w_k\sim N(0,R_k)$. In all numerical computations, the \textbf{R} language (2022) was used. To get the results, a simulation is performed for $K=100$ time-steps with $M=2000$ particles and $N=100$ repeated samples.

\begin{figure}[tp!]
\centering
\hspace{-0.5cm}\includegraphics[width=1\linewidth, height=0.5\textwidth]{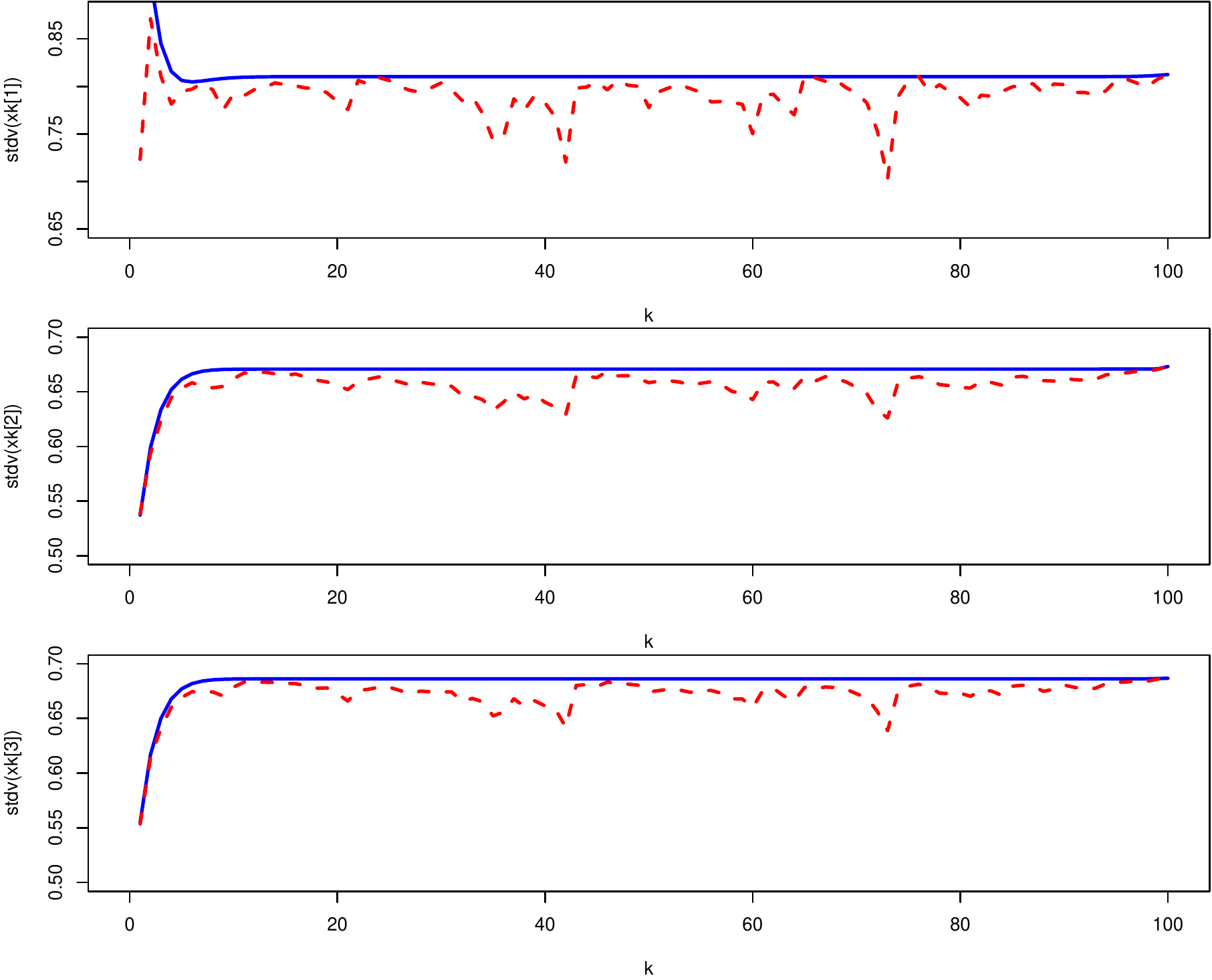}
\caption{Plots of estimated standard errors $\widehat{\sigma}_{k\vert n}$ (dashed) derived from $\sqrt{\widehat{\Sigma}_{x_k,x_k}^s(\theta)}$ (\ref{eq:estvar}) against its theoretical value $\sigma_{k\vert n}$ (solid line) obtained from $\sqrt{\Sigma_{k\vert n}^s}$ (\ref{eq:rtsvar}). The top graph is for $\widehat{\sigma}_{k,1\vert n}$, the second for  $\widehat{\sigma}_{k,2\vert n}$ and the bottom for  $\widehat{\sigma}_{k,3\vert n}$. } \label{fig:stdev}
\end{figure}

Using the complete-data score function (\ref{eq:scoreLN}), the information matrix $J _{x_k,x_k}^z(x_k,x_{k+1},y_{0:n}\vert \theta)$ is given by $H_k^{\top}R_k^{-1} H_k + Q_k^{-1}+ F_{k+1}^{\top} Q_{k+1}^{-1}F_{k+1}>0$. The RTS smoother $\widehat{x}_{k\vert n}^s$ (\ref{eq:eq38}) is compared to (\ref{eq:algo2}) which reads
%%
%\begin{align*}
%&\widehat{x}_{k\vert n}^{s(\ell+1)}=\widehat{x}_{k\vert n}^{s(\ell)} + \big[H_k^{\top}R_k^{-1} H_k + Q_k^{-1}+ F_{k+1}^{\top} Q_{k+1}^{-1}F_{k+1}\big]^{-1}\\
%&\Big(H_k^{\top}R_k^{-1}[y_k-H_k \widehat{x}_{k\vert k}] + F_{k+1}^{\top}Q_{k+1}^{-1}[\widehat{x}_{k+1\vert n}^s -F_{k+1} \widehat{x}_{k\vert k}]\\
%&+ Q_k^{-1}\sum_{m=1}^M w_{k-1}^m(\widehat{x}_{k\vert n}^{s(\ell)}\vert\theta)[F_k x_{k-1}^m -\widehat{x}_{k\vert k}]\Big).
%\end{align*}
%%
%
\begin{align}\label{eq:smcsmoother}
\widehat{x}_{k\vert n}^{s(\ell+1)}=& \big[H_k^{\top}R_k^{-1} H_k + Q_k^{-1}+ F_{k+1}^{\top} Q_{k+1}^{-1}F_{k+1}\big]^{-1} \nonumber\\&\times \big[H_k^{\top}R_k^{-1} y_k+ F_{k+1}^{\top}Q_{k+1}^{-1} \widehat{x}_{k+1\vert n}^s \\&\hspace{0.75cm}+ Q_k^{-1}F_k\sum_{m=1}^M w_{k-1}^m(\widehat{x}_{k\vert n}^{s(\ell)}\vert\theta) x_{k-1}^m \big]. \nonumber
\end{align}
\begin{figure}[tp!]
\centering
\hspace{-0.5cm}\includegraphics[width=1\linewidth, height=0.5\textwidth]{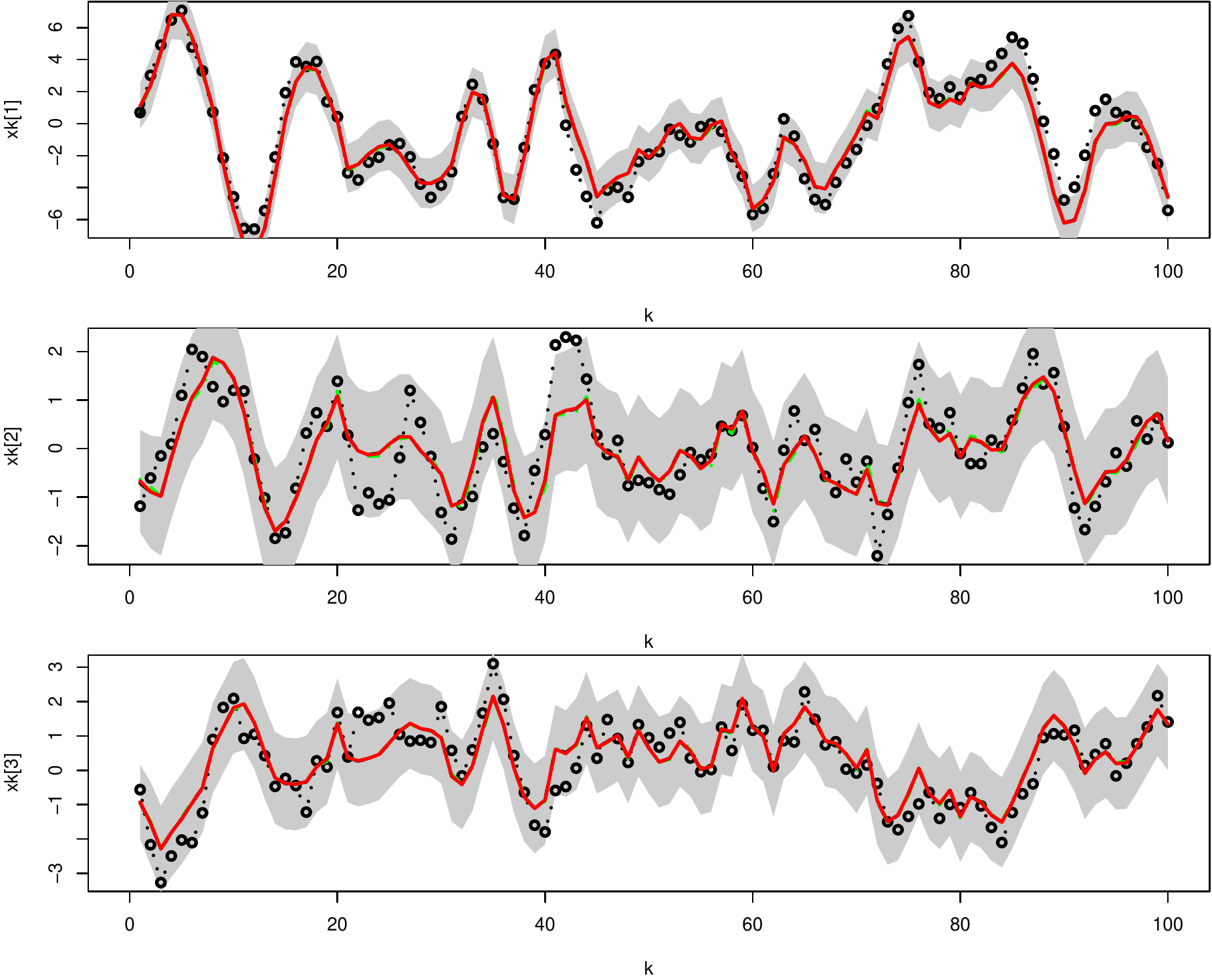}
\caption{Plots of true state $x_k$ ($-o-$), the Kalman filter $\widehat{x}_{k\vert k}$ (dotdashed), RTS smoother $\widehat{x}_{k\vert n}^s$ (\ref{eq:eq38}) (dashed) and the MLE smoother $\widehat{x}_{k\vert n}^s$ obtained using sequential Monte Carlo (\ref{eq:smcsmoother}) (solid) along with $95\%$ confidence interval. The top graph is for $x_{k,1}$, the second for $x_{k,2}$ and the bottom for $x_{k,3}$.} \label{fig:rauch}
\end{figure}
%\begin{enumerate}
%%
%\item[(i)] covariance matrices $\widehat{P}_{20}$, $\Omega_{20}$ and $P_{20\vert 20}$ are resp.,
%%
%\begin{align*}
%\widehat{P}_{20}=&\left(
%\begin{array}{ccc}
%  0.6524 & -0.0857 & 0.0792\\
% -0.0857 &  0.4460 & -0.4105\\
%  0.0792 & -0.4105 & 0.4646
%\end{array}\right),\\[4pt]
%\Omega_{20}=&\left(
%\begin{array}{ccc}
% 0.6524 & -0.0857 &  0.0792\\
% -0.0857 &  0.4460 & -0.4105\\
%  0.0792 & -0.4105 &  0.4646
%\end{array}\right),\\[4pt]
%P_{20\vert 20}=&\left(
%\begin{array}{ccc}
% 0.6601 & -0.0867 & 0.0801\\
% -0.0867 & 0.4530 & -0.4175\\
% 0.0801 & -0.4175 & 0.4716
%\end{array}\right),
%\end{align*}
%\item[(ii)] covariance matrices $\widehat{P}_{53}$, $\Omega_{53}$, and $P_{53\vert 53}$ are 
%%
%\begin{align*}
%\widehat{P}_{53}=&\left(
%\begin{array}{ccc}
% 0.6571 & -0.0858 & 0.0792\\
% -0.0858 & 0.4513 & -0.4159\\
% 0.0792 & -0.4159 & 0.4701
%\end{array}\right),\\[4pt]
%\Omega_{53}=&\left(
%\begin{array}{ccc}
% 0.6571 & -0.0858 & 0.0792\\
% -0.0858 & 0.4513 & -0.4159\\
%  0.0792 & -0.4159 & 0.4701
%\end{array}\right),\\[4pt]
%P_{53\vert 53}=&\left(
%\begin{array}{ccc}
% 0.6601 & -0.0867 &  0.0801 \\
% -0.0867 &  0.4530 & -0.4175\\
%  0.0801 & -0.4175 & 0.4716
%\end{array}\right).
%\end{align*}
%%
%\end{enumerate}
Fig. \ref{fig:stdev} exhibits the standard errors $\widehat{\sigma}_{k\vert n}$ of ML smoother $\widehat{x}_{k\vert n}^s$ (\ref{eq:smcsmoother}) given by the respective diagonal elements of $\sqrt{\widehat{\Sigma}_{x_k,x_k}^s(\theta)}$ (\ref{eq:estvar}) and their theoretical values counterpart $\sigma_{k\vert n}$ derived from $\sqrt{\Sigma_{k\vert n}^s}$ (\ref{eq:rtsvar}). That is $\widehat{\sigma}_{k,1\vert n}^2:=\big[\widehat{\Sigma}_{x_k,x_k}^s(\theta)\big]_{11}$, the first diagonal element of $\widehat{\Sigma}_{x_k,x_k}^s(\theta)$. Similarly defined for $\widehat{\sigma}_{k,2\vert n}^2$ and $\widehat{\sigma}_{k,3\vert n}^2$. Although there are some variations in $\widehat{\sigma}_{k\vert n}$, but overall, the estimates show the convergence to their theoretical values $\sigma_{k\vert n}$, the upper bound. These variations may be reduced by increasing $N$, the number of repeated samples.

Fig. \ref{fig:rauch} displays the smoothing result for the linear system (\ref{eq:eq35})-(\ref{eq:eq36}). Despite relatively high degree of uncertainty in the observation $y_k$ of state-vector $x_k$ through noisy convolution of $x_{k,2}$ and $x_{k,3}$, the Kalman filter $\widehat{x}_{k\vert k}$
%\footnote{defined recursively by $\widehat{x}_{k\vert k}=\widehat{x}_{k\vert k-1} + K_k (y_k - H_k \widehat{x}_{k\vert k-1})$, with $\widehat{x}_{k\vert k-1}= F_k \widehat{x}_{k-1\vert k-1} + G_ku_k$ and $P_{k\vert k-1}=F_kP_{k-1\vert k-1}F_k^{\top} + Q_k$, $\widehat{x}_{0\vert -1}=\mu$ and $P_{0\vert -1}=P_0$}, 
shows similar behavior to the true state $x_k$. In particular, we notice that iterative RTS smoother $\widehat{x}_{k\vert n}^s$ (\ref{eq:eq38}) and ML smoother obtained recursively from (\ref{eq:smcsmoother}) coincides with the Kalman filter $\widehat{x}_{k\vert k}$. The results of Proposition \ref{prop:efficiency} verify the latter findings. To be more details, since the filtering distribution $f(x_k\vert y_{0:k},\theta)=N(x\vert \widehat{x}_{k\vert k},P_{k\vert k}^{-1})$ is symmetry and unimodal, the solution to score equation $\frac{\partial \log f(\widehat{x}_{k\vert n}^s, y_{0:k}\vert\theta)}{\partial x_k}=0$ corresponds to the Kalman filter $\widehat{x}_{k\vert k}$. Moreover, the $95\%$ confidence interval $(\widehat{x}_{k\vert n}^s \pm 1.96 \widehat{\sigma}_{\widehat{x}_{k\vert n}^s})$ for $\widehat{x}_{k,2\vert n}^s$ is wider than those for $\widehat{x}_{k,1\vert n}^s$ and $\widehat{x}_{k,3\vert n}^s$, with $\widehat{x}_{k,1\vert n}^s$ having the narrowest confidence interval. Furthermore, we observe that most (true) states $x_k$ are confined (at least 95 out of 100) in the interval.

%\begin{figure}[tp!]
%\centering
%\includegraphics[width=.865\linewidth]{stdeviter.pdf}
%\caption{Plots of estimated variance $\widehat{\textrm{var}}(\widehat{x}_{k})$, the diagonal element of the recursive covariance estimator $\Omega_k$ (\ref{eq:recOFI}), and that of obtained from the posterior variance $P_{k\vert k}$ (\ref{eq:kalmancovmat}). } \label{fig:stdeviter}
%\end{figure}

%
\begin{figure}[tp!]
\centering
\hspace{-0.5cm}\includegraphics[width=1\linewidth, height=0.5\textwidth]{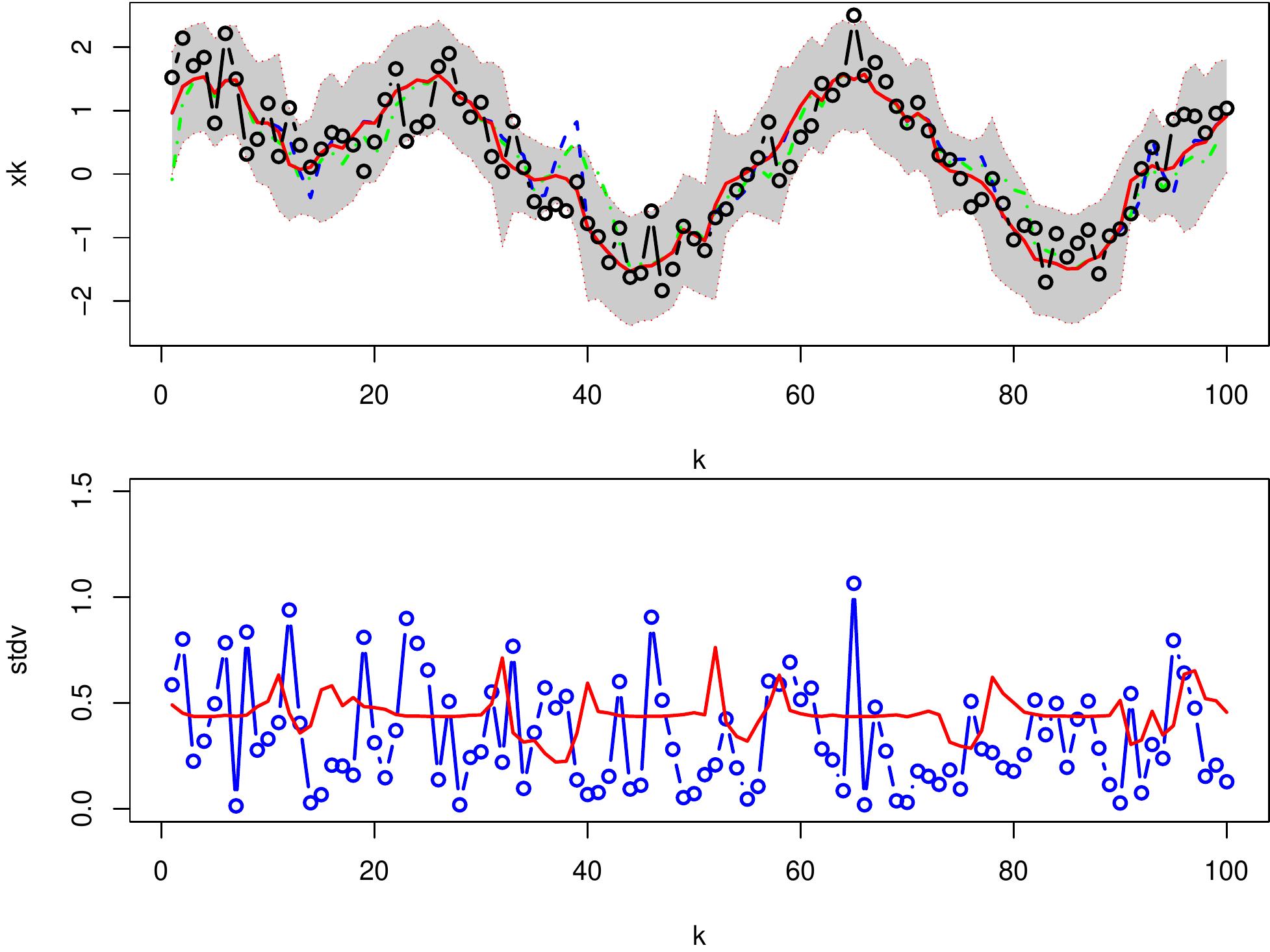}
\caption{Plot of true state $x_k$ ($-o-$), ML state estimator $\widehat{x}_k$ (dashed), and particle filter $\widehat{x}_{k\vert k}$ (dotdashed) and ML smoother $\widehat{x}_{k\vert n}^s$ (solid line). The bottom plot shows the sample standard errors $s_{k\vert n}=\sqrt{\frac{1}{N-1}\sum_{\ell=1}^N (\widehat{x}_{k,\ell\vert n}^s -x_k)^2}$ and their estimated theoretical values $\widehat{\sigma}_{k\vert n}$ derived from $\sqrt{\widehat{\Sigma}_{x_k,x_k}^s}$ (\ref{eq:estvar}).  } \label{fig:smoothernl}
\end{figure}

\subsection{Nonlinear state-space model}

In this example, the recursive EM-gradient-particle filter (\ref{eq:algo2}) is applied to a nonlinear state-space model
\begin{equation}\label{eq:NLSP}
\begin{split}
x_k=& f_k \tanh(\pi x_{k-1}) + v_k, \\
y_k=& \frac{1}{2} x_k + w_k,
\end{split}
\end{equation}
with $f_k=(1+0.5\sin(2\pi k/20))$, $v_k\sim N(0,0.2)$, $w_k\sim N(0,1)$ and $x_0\sim N(0,1)$. This is the same model considered in \cite{Ramadan} with bi-modal and highly skewed filtering distribution $f(x_k\vert y_{0:k},\theta)$. Due to inherent complexity of $F_{k+1}(x_k,u_{k+1})$, the information matrix (\ref{eq:eq40}) and the score function (\ref{eq:scorenl}) are not simplified to derive a compact form of $\widehat{x}_{k\vert n}^{s(\ell+1)}$ from (\ref{eq:algo2}).

Estimation results are presented in Fig. \ref{fig:smoothernl}. From the top graph we observe that the ML smoother $\widehat{x}_{k\vert n}^s$ has more likelihood adherence to the true state $x_k$ compared to ML state-estimator $\widehat{x}_k=\argmax_{x_k} \log f(x_k,y_{0:k}\vert\theta)$, see \cite{Surya2022b}, and particle filter $\widehat{x}_{k\vert k}$ sampled from filtering distribution $f(x_k\vert y_{0:k},\theta)$ using \cite{Kitagawa} algorithm. These numerical outcomes confirm the results of Proposition \ref{prop:efficiency}. Despite the presence of variations in the sample standard errors $s_{k\vert n}=\sqrt{\frac{1}{N-1}\sum_{\ell=1}^N (\widehat{x}_{k,\ell\vert n}^s -x_k)^2}$, they seem to follow similar behavior of their estimated theoretical values $\widehat{\sigma}_{k\vert n}$ derived from $\sqrt{\widehat{\Sigma}_{x_k,x_k}^s}$ (\ref{eq:estvar}). More importantly, we observe that most states $x_k$ are confined in the confidence interval $(\widehat{x}_{k\vert n}^s \pm 1.96 \widehat{\sigma}_{\widehat{x}_{k\vert n}^s})$, at least 95 out of 100.

\section{Concluding remarks}\label{sec:conclusion}

This paper presented a novel method based on an incomplete-information framework for maximum likelihood smoothing estimation of state. The results generalize earlier works of Rauch (1963, et al. 1965) to account for nonlinearity in dynamical systems of the state. The novelty of the approach is central to formulating $(x_k,y_{0:n})$ as an incomplete data of the systems, which is slightly different from the existing methods. Such formulation allows one applying statistical analysis of incomplete data, developed primarily for parameter estimation (see e.g.,\cite{Dempster} and \cite{Little}), to obtain maximum likelihood smoothing estimation of state. The ML smoother is given recursively in terms of the score function and observed information matrix of the incomplete data, which are evaluated using distribution $f(x_k\vert x_{k+1},y_{0:n},\theta)$ and sequential Monte Carlo method. To estimate the standard errors, a recursive equation for the covariance matrix was derived. For the linear system, the method produces the RTS smoother and shows that it is a fully efficient smoother. Numerical study verifies the main results. The new method can be applied for maximum likelihood smoothing estimation in general state-space models. 

%\section*{Acknowledgements}
%
%%The authors are grateful to an anonymous referee  for very helpful comments.
%He acknowledges the Faculty Strategic Research
%Grant No. 20859 of Victoria University of Wellington and hospitality
%provided by the hosts of his visits.

%{\it Conflict of Interest}: None declared.

\appendix

\section{Proof of Theorem \ref{theo:theo2}}
To prove (\ref{eq:louis1}), it suffices to show that $\mathrm{Cov}\big(\frac{\partial \log f(z\vert\theta)}{\partial x_{k+1}},\linebreak\frac{\partial \log f(z\vert \theta)}{\partial x_k}\big\vert \xi,\theta\big)=
\mathbb{E}\big[\frac{\partial \log f(z\vert\theta)}{\partial x_{k+1}}\frac{\partial \log f(z\vert \theta)}{\partial x_k^{\top}}\big\vert \xi,\theta\big]$. Indeed, it holds following (\ref{eq:eq3}) and taking account of $\mathrm{Cov}\big(\frac{\partial \log f(\xi\vert\theta)}{\partial x_{\ell_1}},\frac{\partial \log f(z\vert \xi \theta)}{\partial x_{\ell_2k}}\big\vert \xi,\theta\big)=0,$ $\ell_1,\ell_2\in \{k,k+1\}$, $\ell_1\neq \ell_2$, and $\mathbb{E}\big[\frac{\partial \log f(z\vert \xi,\theta)}{\partial x_{\ell}}\big\vert \xi,\theta\big]=0$, $\ell\in \{k,k+1\}$, by Lemma \ref{lem:lem1}. Using (\ref{eq:eq9}) and (\ref{eq:eq3}) complete the proof. $\square$

\section{Proof of Proposition \ref{prop:prop2}}
Following similar arguments to the proof of Proposition 6 in \cite{Surya2022b}, the boundary condition $f(x_k\vert x_{k-1},\theta)=0$, for $x_k\in\partial \mathcal{X}_k$, results in $\mathbb{E}[\frac{\partial \log f(x_k\vert \theta)}{\partial x_k}\vert \theta]=0$, while the Neumann boundary condition $\frac{\partial f(x_k\vert x_{k-1},\theta)}{\partial x_k}=0$, for $x_k\in \partial \mathcal{X}_k$, leads to $\mathbb{E}\big[-\frac{\partial^2 \log f(x_k\vert\theta)}{\partial x_k \partial x_k^{\top}}\big \vert \theta\big]=\mathbb{E}\big[\frac{\partial \log f(x_k\vert \theta)}{\partial x_k} \frac{\partial \log f(x_k\vert \theta)}{\partial x_k^{\top}}\big\vert \theta\big]$. The claim (\ref{eq:unbiased}) is established using (\ref{eq:mainidentity}) due to $x_k$ considered as an incomplete observation of $(x_k,x_{k+1},y_{0:n})$, i.e., $\frac{\partial \log f(x_k\vert \theta)}{\partial x_k}=\mathbb{E}\big[\frac{\partial \log f(x_k,x_{k+1}, y_{0:n}\vert\theta)}{\partial x_k}\big\vert x_k,\theta\big]$. By the same consideration as the latter, it follows from (\ref{eq:louis1}) that $-\frac{\partial^2 \log f(x_k\vert\theta)}{\partial x_k\partial x_k^{\top}}=\mathbb{E}\big[-\frac{\partial^2 \log f(x_k,x_{k+1},y_{0:n}\vert\theta)}{\partial x_k \partial x_k^{\top}}\big\vert x_k,\theta\big]-\mathbb{E}\big[\frac{\partial \log f(x_k, x_{k+1},y_{0:n}\vert\theta)}{\partial x_k}\frac{\partial \log f(x_k,x_{k+1},y_{0:n}\vert\theta)}{\partial x_k^{\top}}\big\vert x_k,\theta\big]+\frac{\partial \log f(x_k\vert \theta)}{\partial x_k} \frac{\partial \log f(x_k\vert \theta)}{\partial x_k^{\top}}$. The proof of (\ref{eq:louis2}) is complete on account of (\ref{eq:mainidentity}) and taking expectation $\mathbb{E}[\bullet\vert \theta]$ of the above identity. Similarly, the proof for the expectation of outer product of the score function $\mathcal{S}_{x_k}(x_k,x_{k+1}\vert\theta)$ and $\mathcal{S}_{x_{k+1}}(x_k,x_{k+1}\vert\theta)$ is complete using the identity
\begin{align*}
&\frac{\partial^2 \log f(x_k,x_{k+1}\vert \theta)}{\partial x_k \partial x_{k+1}^{\top}}=\frac{1}{f(x_k,x_{k+1}\vert \theta)} \frac{\partial^2 f(x_k,x_{k+1}\vert\theta)}{\partial x_k,\partial x_{k+1}^{\top}}\\
&\hspace{1cm}-\frac{\partial \log f(x_k,x_{k+1}\vert\theta)}{\partial x_k}\frac{\partial \log f(x_k,x_{k+1}\vert\theta)}{x_{k+1}^{\top}},
\end{align*}
which is obtained by applying the chain rule for derivative. Multiplying both sides by probability density $f(x_k,x_{k+1}\vert\theta)$ and integrating w.r.t $d x_{k+1}dx_k$ yields,
\begin{eqnarray*}
&&\hspace{-0.5cm}\mathbb{E}\big[\frac{\partial^2 \log f(x_k,x_{k+1}\vert\theta)}{\partial x_k \partial x_{k+1}^{\top}}\big\vert \theta\big]=\int \int \frac{\partial^2 \log f(x_k,x_{k+1}\vert\theta)}{\partial x_k \partial x_{k+1}^{\top}} d x_k d x_{k+1}\\
&&\hspace{1cm}-\mathbb{E}\big[\frac{\partial \log f(x_k,x_{k+1}\vert\theta)}{\partial x_k}\frac{\partial \log f(x_k,x_{k+1}\vert\theta)}{\partial x_{k+1}^{\top}}\big\vert \theta\big].
\end{eqnarray*}
Since $f(x_k,x_{k+1}\vert\theta)=f(x_{k+1}\vert x_k,\theta)f(x_k\vert\theta)$ by Bayes formula and the fact that under (A2a) and (A2b) $f(x_k\vert\theta)=0$, $x_k\in\partial \mathcal{X}_k$, $\Delta\big(\frac{\partial f(x_{k+1}\vert x_k,\theta)}{\partial x_{k+1}}f(x_k\vert\theta)\big\vert_{x_k\in\partial \mathcal{X}_k}=0$ implying $\int \int \frac{\partial^2 \log f(x_k,x_{k+1}\vert\theta)}{\partial x_k \partial x_{k+1}^{\top}} d x_k d x_{k+1}=0$. Hence,
\begin{align*}
&\mathbb{E}\big[\frac{\partial^2 \log f(x_k,x_{k+1}\vert\theta)}{\partial x_k \partial x_{k+1}^{\top}}\big\vert\theta\big]=\\
&\hspace{1cm}-\mathbb{E}\big[\frac{\partial \log f(x_k,x_{k+1}\vert\theta)}{\partial x_k}\frac{\partial \log f(x_k,x_{k+1}\vert\theta)}{\partial x_{k+1}^{\top}}\big\vert\theta\big].
\end{align*}
Using $\xi:=(x_k,x_{k+1})$ as an incomplete observation of $z:=(x_k,x_{k+1},y_{0:n})$, the proof is complete using (\ref{eq:louis2}). The proof for the second row of the matrix identity (\ref{eq:eq14}) can be established in similar way to the first row. $\square$

\section{Proof of Proposition \ref{prop:efficiency}}
Applying the identity (\ref{eq:mainidentity}) by setting $\xi:=(\widehat{x}_{k\vert n}^s,y_{0:k})$ as an incomplete observation of $z=(\widehat{x}_{k\vert n}^s, \widehat{x}_{k+1\vert n}^s, y_{0:n})$ we obtain after taking expectation $\mathbb{E}[\bullet\vert \widehat{x}_{k\vert n}^s,y_{0:k},\theta]$ on both side of (\ref{eq:eq20}), $\mathbb{E}\big[\frac{\partial \log f(\widehat{x}_{k\vert n}^s, \widehat{x}_{k+1\vert n}^s,y_{0:n}\vert\theta_k)}{\partial x_k}\big\vert \widehat{x}_{k\vert n}^s,y_{0:k},\theta]=\frac{\partial \log f(\widehat{x}_{k\vert n}^s, y_{0:k} \vert \theta)}{\partial x_k}=0,$ leading to $
2\big[ \log f(\widehat{x}_{k\vert n}^s, y_{0:k}\vert\theta)-\log f(x_k^0,y_{0:k}\vert\theta)\big]\frac{\partial \log f(\widehat{x}_{k\vert n}^s, y_{0:k}\vert \theta)}{\partial x_k}=0,$
which proves the first claim. To establish the second claim, it was shown in VanTrees (1968, et al. 2013), see also \cite{Surya2022b}, that the ML estimator $\widehat{x}_k$ has the covariance matrix $\mathbb{E}[(\widehat{x}_k-x_k)(\widehat{x}_k-x_k)^{\top}\vert \theta]$ given by the inverse of expected information matrix $I(\theta)=\mathbb{E}[ -\frac{\partial^2 \log f(x_k,y_{0:k}\vert\theta)}{\partial x_k \partial x_k^{\top}}\vert \theta]$.  It remains to show that $\Sigma_{x_k,x_k}^s(\theta)\leq I^{-1}(\theta)$. On account that $\frac{\partial \log f(\widehat{x}_k,y_{0:k}\vert\theta)}{\partial x_k}=0$, the proof follows from applying identities (\ref{eq:louis1}) of Theorem \ref{theo:theo2} and (\ref{eq:mainidentity}) for the vectors $\xi=(\widehat{x}_k,y_{0:k})$ and $z=(\widehat{x}_k,\widehat{x}_{k\vert n+1}^s, y_{0:n})$ which now reads 
\begin{align*}
&-\frac{\partial^2 \log f(\widehat{x}_k,y_{0:k}\vert\theta)}{\partial x_k\partial x_k^{\top}}\\
&=\mathbb{E}\Big[-\frac{\partial^2 \log f(\widehat{x}_k, \widehat{x}_{k+1\vert n}^s, y_{0:n}\vert\theta)}{\partial x_k\partial x_k^{\top}}\Big\vert \widehat{x}_k, y_{0:k},\theta\Big]\\
&+\mathbb{E}\Big[\frac{\partial \log f(\widehat{x}_k, \widehat{x}_{k+1\vert n}^s, y_{0:n}\vert\theta)}{\partial x_k}\Big\vert \widehat{x}_k, y_{0:k},\theta\Big]\\
&\hspace{1cm}\times \mathbb{E}\Big[\frac{\partial \log f(\widehat{x}_k, \widehat{x}_{k+1\vert n}^s, y_{0:n}\vert\theta)}{\partial x_k^{\top}}\Big\vert \widehat{x}_k, y_{0:k},\theta\Big]\\
&\hspace{-0.5cm}-\mathbb{E}\Big[\frac{\partial \log f(\widehat{x}_k, \widehat{x}_{k+1\vert n}^s, y_{0:n}\vert\theta)}{\partial x_k}  \frac{\partial \log f(\widehat{x}_k, \widehat{x}_{k+1\vert n}^s, y_{0:n}\vert\theta)}{\partial x_k^{\top}}\Big\vert \widehat{x}_k, y_{0:k},\theta\Big].
\end{align*}
Note that since $\widehat{x}_{k}$ is the maximizer of $\log f(x_k,y_{0:k}\vert\theta)$, $\frac{\partial \log f(\widehat{x}_k, \widehat{x}_{k+1\vert n}^s, y_{0:n}\vert\theta)}{\partial x_k}\neq 0$. However, on account that $\mathbb{E}\big[\frac{\partial \log f(\widehat{x}_k, \widehat{x}_{k+1\vert n}^s, y_{0:n}\vert\theta)}{\partial x_k}\big\vert \widehat{x}_k, y_{0:k},\theta\big]=\frac{\partial \log f(\widehat{x}_k,y_{0:k}\vert\theta)}{\partial x_k}$, see identity (\ref{eq:mainidentity}), the outer product of conditional expectations are therefore equal to zero. It implies a.s. that  
\begin{align*}
&\mathbb{E}\Big[-\frac{\partial^2 \log f(\widehat{x}_k, \widehat{x}_{k+1\vert n}^s, y_{0:n}\vert\theta)}{\partial x_k\partial x_k^{\top}}\Big\vert \widehat{x}_k, y_{0:k},\theta\Big]\\
&\hspace{1cm} >-\frac{\partial^2 \log f(\widehat{x}_k,y_{0:k}\vert\theta)}{\partial x_k\partial x_k^{\top}}>0,
\end{align*}
which for given observations $y_{0:k}$ holds for all estimates $\widehat{x}_k$. This inequality agrees with (\ref{eq:eq10}). It results in 
\begin{align*}
&\mathbb{E}\Big[-\frac{\partial^2 \log f(\widehat{x}_k, \widehat{x}_{k+1\vert n}^s, y_{0:n}\vert\theta)}{\partial x_k\partial x_k^{\top}}\Big\vert \widehat{x}_k, y_{0:k},\theta\Big]>\\
&\mathbb{E}\Big[\frac{\partial \log f(\widehat{x}_k, \widehat{x}_{k+1\vert n}^s, y_{0:n}\vert\theta)}{\partial x_k}  \frac{\partial \log f(\widehat{x}_k, \widehat{x}_{k+1\vert n}^s, y_{0:n}\vert\theta)}{\partial x_k^{\top}}\Big\vert \widehat{x}_k, y_{0:k},\theta\Big],
\end{align*}
which for given observations $y_{0:k}$ holds for all $\widehat{x}_k$.
Using the two inequalities, the proof follows from (\ref{eq:eq22}). $\square$

\medskip

\begin{wrapfigure}{l}{20mm}
    \includegraphics[width=1in,height=1.25in,clip,keepaspectratio]{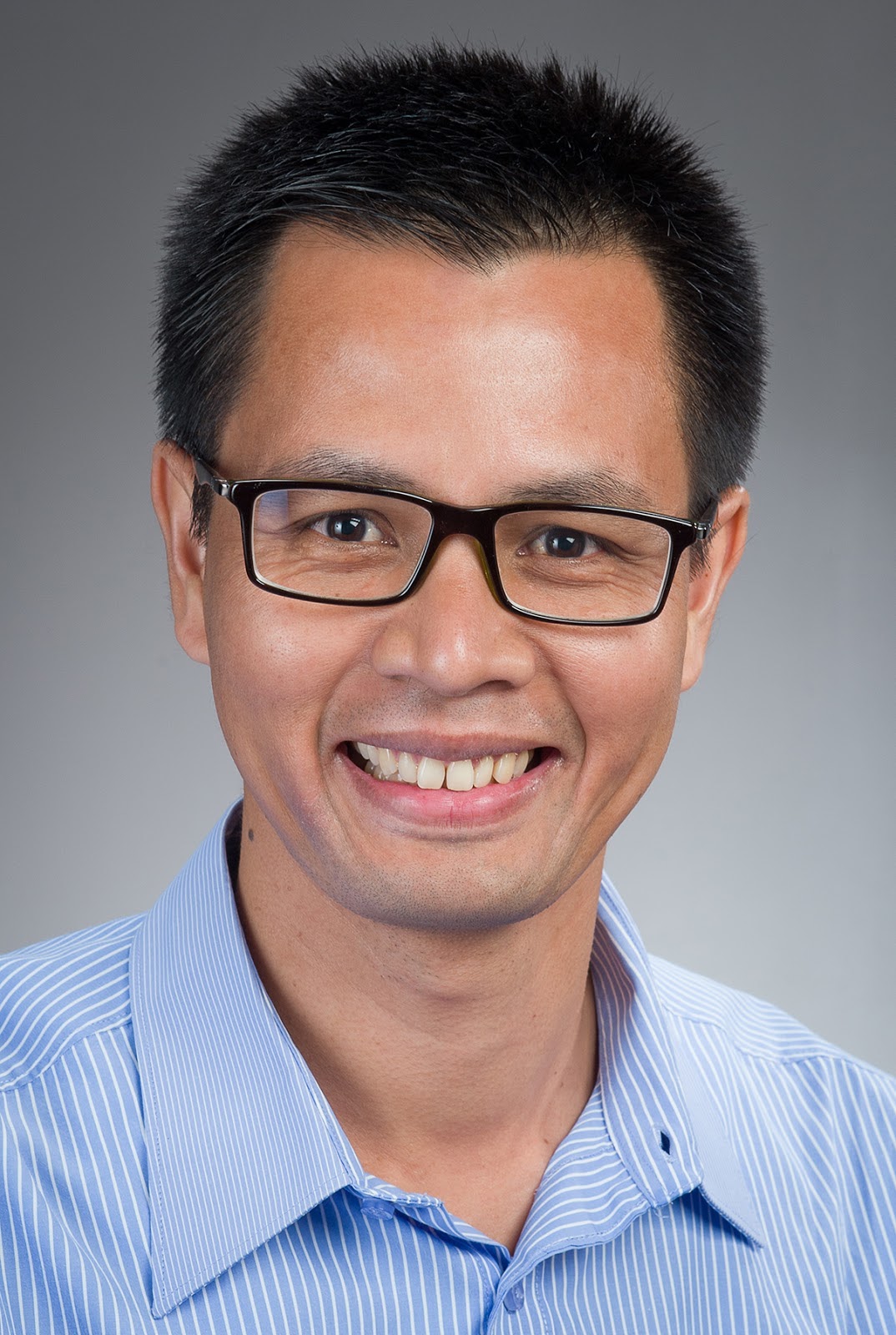}
  \end{wrapfigure}\par
  \textbf{Budhi Arta Surya} is a senior lecturer (equivalent to an associate professor) of statistics at the School of Mathematics and Statistics of Victoria University of Wellington, New Zealand. He received an ingenieur Ir. (equivalent to MSc) degree in applied mathematics (majoring in stochastic systems and control) from the University of Twente, Enschede, in August 2001 and a doctorate degree Dr. in applied mathematics (applied probability and stochastic process) in January 2007 from the University of Utrecht, both in Netherlands. His doctorate thesis was concerned with optimal stopping of L\'evy processes. Soon after graduating from Utrecht, he joined  Bank of America Corp. as a quantitative financial analyst, based mainly in Singapore, with direct reporting line to the Head of Quantitative Risk Management Group in Charlotte, USA. He held DAAD Visiting Scholarship to Mathematical Institute of Goethe University of Frankfurt in October-November 2013 and was visiting scholar to Department of Statistics of London School of Economics in October 2013 and February 2018, to Department of Industrial Engineering and Operations Research of Columbia University in May 2014 and to Department of Technology, Operations and Statistics of the Stern School of Business of New York University in September 2019.  \par
  
  \end{document}